\documentclass{ws-rv9x6}	%	\def\draftmode{}
\usepackage{ws-rv-van}             % numbered citation/references (default)
\makeindex
\usepackage{bbm}	\def\Id{{\mathbbm 1}}	\let\id=\Id      

\def\ifundefined#1{\expandafter\ifx\csname#1\endcsname\relax}
\ifundefined{draftmode}	\else	\input draftmode	\fi

\def\NewTitle{		Heterotic (0,2) Gepner Models and Related Geometries}
\def\NewAbstract{
	On the sad occasion of contributing to the memorial volume
	``Fundamental Interactions'' for my teacher Wolfgang Kummer
	I decided to recollect and extend some unpublished notes from the 
	mid 90s when I started to build up a string theory group in Vienna 
	under Wolfgang as head of the particle physics group. His extremely
	supportive attitude was best expressed by his saying that one should
	let all flowers flourish.
	I hope that these notes will be useful in particular in view of 
	the current renewed interest in heterotic model building.

	The content of this contribution is based on the bridge between
	exact CFT and geometric techniques that is provided by the
	orbifold interpretation of simple current modular invariants.
	After reformulating the Gepner construction in this language I
	describe the generalization to heterotic (0,2) models and its
	application to the Geometry/CFT equivalence between Gepner-type 
	and Distler-Kachru models that was proposed by Blumenhagen, 
	Schimmrigk and Wisskirchen. We analyze a series of solutions to 
	the anomaly equations, discuss the issue of mirror symmetry, 
	and use the extended Poincar\'e polynomial to extend the construction 
	to Landau-Ginzburg models beyond the realm of rational CFTs.
	In the appendix we discuss Gepner points in torus orbifolds, 
	which provide further relations to free 
	bosons and free fermions, as well as %mirror constructions.
	simple currents in $N=2$ SCFTs and minimal models.
}

\let\a=\alpha   \let\b=\beta    \let\g=\gamma   \let\d=\delta   %% --> GREEK
         \let\th=\theta  \let\e=\varepsilon
   \let\l=\lambda  \let\m=\mu      
      \let\x=\xi      \let\p=\pi      \let\r=\rho     \let\s=\sigma 
\let\t=\tau        \let\c=\chi     \let\ps=\psi    
\let\ph=\varphi \let\Ph=\phi	\let\Ps=\Psi       \let\S=\Sigma 
\let\P=\Pi             \let\D=\Delta

  \def\IP{{\Bbb P}} \def\IW{{\Bbb W}}
 \def\IN{{\Bbb N}} \def\IZ{{\Bbb Z}}  

\def\EEL#1 {\label{#1}\EE}      \let\nn=\nonumber       %% --> EQUATION macros
\def\BE {\begin{equation}}      \def\EE {\end{equation}}        
\def\BEA{\begin{eqnarray}}      \def\EEA{\end{eqnarray}} 
\def\BI {\begin{itemize}}       \def\EI {\end{itemize}}        

\def\mao#1{\mathop{\rm #1}\nolimits}   % amssymb: \pmb \frac \text \boldsymbol
\def\tr{\mao{tr}}              
\def\mod{\mao{mod}}     \def\gcd{\mao{gcd}}     

\let\and=\wedge         \let\eq=\equiv          \let\ex=\times
\let\then=\Rightarrow      
\let\hc=\dagger                
\let\bra=\langle        \let\ket=\rangle        \def\<#1\>{\bra #1 \ket}

\let\0=\over      \let\1=\vec      \def\2{{1\over2}}    \let\3=\ss
\let\4=\underline \let\5=\overline \let\6=\partial      \def\7#1{{#1}\llap{/}}
\def\8#1{{\textstyle{#1}}}         \def\9#1{{\ifmmode{\pmb{#1}}\else\bf#1\fi}}

\def\ca{{\cal A}}  \def\cc{{\cal C}} 
 \def\cf{{\cal F}} \def\cg{{\cal G}} \def\ch{{\cal H}}
   
 \def\cn{{\cal N}} \def\co{{\cal O}} \def\cp{{\cal P}} 
   \def\ct{{\cal T}} 
\def\cu{{\cal U}}    
  
					\def\rel#1 #2{\buildrel #1 \over {#2}}

\def\HS#1 {\hspace*{#1pt}} \def\VS#1 {\vspace*{#1pt}} \long\def\del#1\enddel{}
\def\BC{\begin{center}}   \def\VR#1#2{\vrule height #1mm depth #2mm width 0pt}
\def\EC{\end{center}}      \def\TVR#1#2{@{~~\VR{#1}{#2}}}

\def\su{{v}}	\def\sp{{s}}	\def\sc{{\bar s}} 	\def\X{{C}}
\def\Jsp{{J_\sp}}		\def\GSO{{GSO}} 	\def\JGSO{J_{GSO}}
\let\fns=\footnotesize	\let\ns=\normalsize		\newcounter{MyTable}
   
\def\plb#1 #2 {Phys. Lett. {\bf B#1} #2 }
\def\phr#1 #2 {Phys. Rep. {\bf  #1} #2 }        
\def\npb#1 #2 {Nucl. Phys. {\bf B#1} #2 }
\def\aph#1 #2 {Ann. Phys. {\bf #1} #2 }         
\def\jmp#1 #2 {J. Math. Phys. {\bf #1} #2 }
\def\jgp#1 #2 {J. Geom. Phys. {\bf #1} #2 }
\def\prd#1 #2 {Phys. Rev. {\bf D#1} #2 }
\def\prl#1 #2 {Phys. Rev. Lett. {\bf #1} #2 }
\def\rmp#1 #2 {Rev. Mod. Phys.  {\bf #1} #2 }
\def\zpc#1 {Z. Phys. {\bf #1C} }
\def\cmp#1 #2 {Commun. Math. Phys. {\bf #1} #2 }
\def\cqg#1 #2 {Class.Quant.Grav. {\bf #1} #2 }
\def\mpl#1 {Mod. Phys. Lett. {\bf A#1} }
\def\cpc#1 {Computer Phys. Commun. {\bf #1} }   % Belfast,cpc@v1.am.qub.ac.uk
\def\ijmp#1 {Int. J. Mod. Phys. {\bf A#1} }
\def\ijmpC#1 {Int. J. Mod. Phys. {\bf C#1} }

\def\BP{\begin{picture}} \def\EP{\end{picture}}         %% --> PICTURE macros 
\def\putlab#1)#2#3{\put#1){\makebox(0,0)[#2]{\small #3}}}
\def\putlin#1,#2,#3,#4,#5){\put#1,#2){\line(#3,#4){#5}}} %\putlin(x,y,dx,dy,l)
\def\putvec#1,#2,#3,#4,#5){\put#1,#2){\vector(#3,#4){#5}}}
\def\putc#1)#2{\put#1){\makebox(0,0)[c]{#2}}}

		\long\def\old#1\endold{{\bf #1}}
		\long\def\new#1\endnew{{\bf #1}}

						\long\def\Count#1\EndCount{#1}
						\long\def\Count#1\EndCount{}

\begin{document}		\chapter{\NewTitle}

\author[Maximilian KREUZER]{Maximilian KREUZER\footnote{Email:
	\texttt{Maximilian.Kreuzer@tuwien.ac.at}}}
\address{%\footnotemark[0]
       Institute for Theoretical Physics, Vienna University of Technology\\
       Wiedner Hauptstrasse 8--10, A-1040 Vienna, AUSTRIA}

\begin{abstract}	\NewAbstract	\end{abstract}		\body

\del
\vspace{-9pt}\vbox to 0pt{\tiny\vspace{40mm}
\begin{tabbing}
Introduction: \=	Gepner and WP spaces: LG and GLSM\\
	\>Mirror Symmetry: Greene-Plesser, BH and Batyrev\\
\>CFT: simple currents and orbifolds, proof of BH, include DT (not Z2)\\
\>Contents
\\2. Orbifolds, simple currents and quantum symmetries:
	SCMI and orbifolds with discrete torsion
\\3. Gepner-type (0,2) models:
	Bosonic string map and generalized GSO projection,
	Extensions, gauge groups and (0,2) models
\\4. Extended Poincare Polynomials and mirror symmetry
\\5. Geometry/CFT: 
	Weighted projective and vector bundle data
\\6. Perspectives
\\Appendix A: Gepner models and Torus orbifolds
\\Appendix B: Simple currents in N=2 SCFT minimal models
\\Appendix C: Toric resolutions
\end{tabbing}}
\del\enddel

\section{Introduction}

When a number of differenent constructions for heterotic string 
compactifications % model building 
were developed in the late 1980s it soom became clear from the coincidence 
of spectra that Gepner models\cite{Gepner} 
%, built from N=2 superconformal minimal models, 
and Calabi-Yau hypersurfaces in weighted projective spaces 
\cite{CLS} should be closely related.
The connection was found to be provided by Landau-Ginzburg models \cite{va89},
whose superpotential $W(\Ph_i)$ can be identified with the hypersurface 
equation $W(z_i)=0$. A Fermat-type potential of the form $W=\sum \Ph_i^{K_i}$ 
then corresponds to a Gepner model with levels $k_i=K_i-2$. 
The precise relation was later derived by Witten by virtue of
his $N=2$ supersymmetric
gauged linear sigma model (GLSM)\cite{wi93}, which --
in addition to the shape parameters (complex structure moduli) 
%% of the Calabi-Yau hypersurface that are encoded 
in the superpotential $W$ -- contains the size parameters 
(K\"ahler moduli) of the Calabi-Yau % hypersurface
as D-terms.

The Gepner point thus turns out to be located at small values of the
K\"ahler moduli, % where $\L$ obtains a VEV, 
way outside the range of
validity of sigma model perturbation theory, so that
Gepner models provide an exactly solvable CFT stronghold inmidst the
realm where strong quantum corrections invalidate any naive geometrical
picture. This proved to be useful in many contexts like closed string
mirror symmetry\cite{MS}, as well as homological mirror symmetry, where,
for example, the transport of exact CFT boundary states to D-branes at large 
volume can be studies \cite{HHP}.

In the context of perturbative heterotic strings the phenomenological 
condition of space-time supersymmetry in the RNS formalism implies that 
the (0,1) superconformal invariance that is left over from the gauge-fixed 
world-sheet %%right-moving 
supergravity%
\footnote{~	The (0,1) superconformal invariance is hence required for a 
		consistent coupling to the superghosts in BRST quantization.
} 
is extended to a (0,2) superconformal invariance plus 
quantization of the U(1) charges.\cite{hu85,ba88} This is, in fact, an
equivalence, because quantization of the N=2 superconformal U(1) charge
implies locality of the spectral flow operator, which implements the 
space-time SUSY transformations on the internal CFT part of vertex 
operators.\cite{ba88}

In the geometric context (0,2) models correspond to stable holomorphic
vector bundles $V_1\times V_2\subset E_8\times E_8$ on a Calabi-Yau 
manifold $X$ with vanishing first Chern classes satisfying the anomaly 
cancellation condition
$ch_2(V_1)+ch_2(V_2)=ch_2(TX)$. The notion of a (2,2) model then refers to the 
choice $V_1=TX$ with trivial $V_2$, called standard embedding,
so that the structure group $SU(3)$ of $TX$ breaks $E_8\times E_8$ to the gauge
group $E_6\times E_8$ in 4 dimensions. The name (2,2) originates from the
CFT analog of this situation where we replace the compactification manifold
by an abstract $N=(2,2)$ left-right symmetric superconformal field theory
with central charge $c=9$. This ``internal sector'' is
combined with 
%%%	the right-moving space-time superfields and the left-moving bosonic 
%%%	space-time coordinates, 
the 4 space-time coordinates $X^\mu$ and their right-moving superpartners 
$\bar\ps^\mu(\bar z)$,
augmented by a left-moving
$SO(10)\times E_8$ current algebra, whose central charge 13 adds up with
4 non-compact dimensions and the internal $c=9$ to the critical dimension 26 
of the bosonic string.%
\footnote{~
	The value $c=9$ corresponds to 6 compact dimensions $X^i$ plus the
	contribution from their right-moving fermionic superpartners $\ps^i$.
\del
	For K3 compactifications to six dimensions we would have more
	supersymmetry and hence need a $c=6$ $N=(4,4)$ internal SCFT that
	is augmented by a manifest $D_6\times E_8$ current algebra in the 
	bosonic sector. The R-symmetry of the $N=4$ algebra now $SU(2)$,
	instead of $U(1)$, and the bosonic analog of the mechanism that
	creates space-time SUSY extends $D_6\times A_1$ to a space-time
	$E_7$ gauge symmetry.
\enddel
}
The same spectral flow mechanism that generates 
space-time SUSY in the right-moving sector then extends the manifest 
$SO(10)$ times the $U(1)$ of the $N=2$ superconformal algebra to the low 
energy $E_6$ gauge symmetry of the standard embedding.%
\footnote{~
	More precisely, the mechanisms are mapped to one another by the
	bosonic string map and its inverse, the
	Gepner map, respectively (see below).
} 
In the geometric context this amounts to the GSO projection. For a general 
internal $N=2$ SCFT with fractional charges it has to be augmented by charge 
quantization and is then refered to as ``generalized GSO projection''.

% It has been known for a long time that the generic (perturbative)
% supersymmetric string vacua have (0,2) supersymmetry on the world sheet
% \cite{hu85,ba88}.

While the general (0,2) models have better phenomenological prospects,
like featuring the more relalistic GUT gauge groups $SO(10)$ and $SU(5)$,
the (2,2) case has been studied much more systematically.
In the realm of $\s$ models one reason for this was the discovery of world
sheet instanton corrections \cite{we86,di86}, which were believed to 
destabilize the vacua.
%, while they are forbidden by non-renormalization theorems for (2,2).
A criterion for avoiding this problem was soon found by Distler and Greene
\cite{di881}; see also \cite{di871,cv87,gr90}. The technical difficulty of
checking the `splitting type' of the stable vector bundles, however, 
provided a powerful deterrent for further progress. 
The situation became much more secure with
Witten's gauged linear sigma models \cite{wi93}, the (0,2) version of which
was used by Distler and Kachru \cite{di94,ka95} to generalize the construction
introduced by Distler and Greene \cite{di881}. The resulting class of models 
is now
believed to define honest (0,2) SCFTs at the infrared fixed point \cite{si95}.
Somewhat ironically, with the recognition of the importance of moduli 
stabilization for model building, world-sheet instantons can turn 
from an obstacle	%, in a sense, from a hazard 
into a virtue, and one now has to work quite hard \cite{BKOS}
to circumvent the cancellation mechanim that has been established
for toric Calabi-Yau complete intersections by Beasley and Witten \cite{BeWi}.
There is also much recent work on generalizations like heterotic M-theory 
\cite{DLOV} and heterotic compactification with H-field background flux 
\cite{BBFTY}, but this is beyond the scope of the present note.

In the realm of exact methods a powerful generalization of 
Gepner's construction \cite{Gepner} was found by Schellekens and Yankielowicz 
\cite{sc90}, who used simple currents \cite{sc90r} to produce a telephone 
book of (1,2) models \cite{sc90T} from tensor products of minimal models.
For the (0,2) case their huge list of models apparently
was so far from complete that it never was published. At the same time
closely related methods were used by Font et al. \cite{fo89,fo90}
to construct pseudo-realistic models. On the CFT side the main problem
is the arbitrariness in the selection of a reasonable subset from the huge
set of available models. A landmark in this effort was Schellekens' theorem
on the conditions for the possibility of avoiding fraction electric charges
\cite{sc90e,ly96}.

An interesting question is, of course, to what extent the geometric
and the CFT approaches to (0,2) models overlap. 
The identification of models that are accessible to both
constructions would provide further evidence for the stability of the $\s$
model constructions, but most importantly allows to explore deformations 
of the rational models, which only live at certain points in moduli space. 
Originally based on a stochastic computer search for matching particle spectra
Blumenhagen et al. \cite{bl95,bl96} proposed a set of
gauge bundle data on a complete intersection that is conjectured 
to describe the moduli space of a rational superconformal (0,2) cousin of
the Gepner model on the quintic.
Using the classification of simple current modular invariants \cite{kr94}
the product invariant that these authors employ can be translated into the
canonical form \cite{kr96} that exhibits its relation to orbifold twists and 
discrete torsion \cite{va86}.
It turns out that the breaking of the gauge group from $E_6$ to $SO(10)$
is due to a certain twist of order 4 that acts on a minimal model factor 
of the internal conformal field theory (at odd level) and on an $SO(2)$ 
that is part of the linearly realized $SO(2)\ex SO(8)\subset SO(10)$ gauge
symmetry \cite{kr96}.
% Later this was extended to a series of identifications \cite{bl97,kr96}
%
% Based on an analysis of the anomaly matching conditions and on the 
Assuming that the $\IZ_4$ breaking mechanism does not care about the rest of 
the conformal field theory and only acts on a Fermat factor of a 
non-degenerate potential we analysed the anomaly matching conditions
and proposed a whole series of identifications \cite{kr96}
that provides us with 3219 models, based on the list of 7555 weights for 
transverse hypersurfaces in weighted projective spaces \cite{nms,kl94}, 
and many more if we combine this with other constructions like orbifolding
and discrete torsion \cite{aas,ade,lgt}. 

The purpose of this note is to 
collect the necessary ingredients for these constructions, where the 
concept of the extended Poincar\'e polynomial \cite{kr95} is used to 
generalize the CFT approach to Landau-Ginzburg models beyond the exactly 
solvable case.
In section \ref{sec:2} we discuss simple current modular invariants (SCMI)
\cite{sc90r} and their geometric interpretation \cite{kr94}. 
To set up the concepts we begin with recalling the
geometric orbifolding idea and use it to motivate and interpret 
the formula for the most general invariant.
In section \ref{sec:3} 
we use simple current techniques for the implementation
of the generalized GSO projection and show how the Gepner construction
generalizes to (0,2) models in general and, in particular, for gauge symmetry 
breaking in the proposed $\s$ model connection. 
%In section \ref{sec:4} 
We discuss the 
counting of non-singlet spectra in terms of the information encoded in the
extended Poincar\'e polynomial, thus extending the scope of the construction
to arbitrary Landau-Ginzburg orbifolds. Since the charge conjugation of 
$N=2$ minimal models is a simple current modular invariant, our
discussion explains
the observed (0,2) mirror symmetry \cite{bl961,bl971} along the lines of 
Greene-Plesser orbifolds and their generalization due to Berglund and H\"ubsch
\cite{Berglund:1991pp}, which applies to the large class of transversal 
potentials that are minimal in a certain sense \cite{Kreuzer:1994uc}.
Section \ref{sec:5} % is devoted to
briefly recollects the geometric side of the proposed identifications.
Here we start with an ansatz for the base manifold and vector bundle
data that are conjectured to describe the moduli spaces of the (0,2) models
and find a unique solution to the anomaly equations. 
%% For more details
% Here we collect the vector bundle data and present the solutions to the
% anomaly equations, but refer to the relevant literatur for more details
%% on the underlying tools we refer to the relevant literature.
In section \ref{sec:6} we conclude with % by recollecting 
a number of topics for generalizations and further studies.

Some technical points are discussed in appendices. In appendix A we
%% consider relations to Free Boson and Free Fermion constructions. We 
discuss Gepner points of torus orbifolds and %provide %, in particular,
exact CFT realizations for the extensions of $\IZ_2\times\IZ_2$ 
orbifolds recently classified by Donagi and Wendland \cite{DoWe}.
Appendix B discusses simple currents in $N=2$ SCFTs and their use for
explaining %the 
labels and field identifications of $N=2$ minimal models. 
%Appendix C
%briefly discusses the resolution of singularites of the base manifolds 
%used in the geometry correspondence in terms of toric Calabi-Yau
%complete intersections \cite{di96}.

\del
\vspace{-119pt}\vbox to 175pt{\tiny\vspace{40mm}
\begin{tabbing}
Introduction: \=	Gepner and WP spaces: LG and GLSM\\
	\>Mirror Symmetry: Greene-Plesser, BH and Batyrev\\
\>CFT: simple currents and orbifolds, proof of BH, include DT (not Z2)\\
\>Contents
\\2. Orbifolds, simple currents and quantum symmetries:
	SCMI and orbifolds with discrete torsion
\\3. Gepner-type (0,2) models:
	Bosonic string map and generalized GSO projection,
	Extensions, gauge groups and (0,2) models
\\4. Extended Poincare Polynomials and mirror symmetry
\\5. Geometry/CFT: 
	Weighted projective and vector bundle data
\\6. Perspectives
\\Appendix A: Gepner models and Torus orbifolds
\\Appendix B: Simple currents in N=2 SCFT minimal models
\\Appendix C: Toric resolutions
\end{tabbing}}
\del\enddel

\del
Section 2: Orbifolds, simple currents and quantum symmetries:
	SCMI and orbifolds with discrete torsion
\\Section 3. Gepner-type (0,2) models:
	Bosonic string map and generalized GSO projection,
	Extensions, gauge groups and (0,2) models
\\Section 4. Extended Poincare Polynomials and mirror symmetry
\\Section 5. Geometry/CFT: 
	Weighted projective and vector bundle data
\\Section 6. Perspectives
\\Appendix A: Gepner models, Torus orbifolds and Mirror Symmetry
\\Appendix B: Simple currents in N=2 SCFT minimal models
\\Appendix C: Toric resolutions
\enddel

\def\addrefs{

	Monads: 0 - V -f- B -g- C - 0, g a matrix of polynomials and
		B,C direct sums of line bundles $B=\oplus_{i=1}^{r_B}\co(b_i)$
	Friedman-Morgan-Witten: spectral cover on elliptically fibered CYs.

}

\section{Orbifolds and simple currents %and quantum symmetries
							}\label{sec:2}

\def\tbc#1#2{\hbox{\;$#1$}                      % torus boundary conditions
    \hbox to 7mm{\hfil\raisebox{.5mm}{\framebox[5mm]{\rule{0mm}{1mm}}}\hfil}
    \llap{\hbox to 6mm{\hss\raisebox{-3.9mm}{$#2$}\hss}}}

The concept of an orbifold CFT originates from the geometric picture of
closed strings on orbit spaces $X/G$ where $X$ is a smooth manifold with
a discrete group action $G$, with or without fixed points. The modding
out of $G$ has two consequences: String states on the orbifold need to 
be invariant under the symmetry on the covering space $X$, which leads to a 
projection of the Hilbert space $\ch_X$ on the covering space to $G$-invariant 
states. On the other hand, new closed string
states emerge, whose $2\p$-periodicity on $X/G$ corresponds to periodicity up
to a group transformation $g\in G$ on $X$. The Hilbert space has hence to
be augmented by twisted sectors $\ch_X^{(g)}$.

\subsection{Orbifold CFT and modular invariance}
For abstract conformal field theories $\cc$ that are invariant under a group 
$G$ of symmetry transformations the same result can be derived from modular 
invariance and factorization constraints on the partition function without
relying on a geometric interpretation. Depicting the one-loop partition 
function by a torus that indicates the double-periodic boundary conditions
imposed in the path integral,
$
	Z_\cc = \tbc {\!\!} {}	\VS-4
$, 
the orbifold partition function can be obtained as a linear combination
of partition functions with boundary conditions twisted by group 
transformations $g$ and $h$,
\BE
	Z_\cc(g,h) = \tbc gh
\EE
in the vertical and horizontal direction, respectively. If we interpret the
% periodic 
horizontal direction as the spacial extension of a 
closed string and the vertical direction as Euclidean time then $h$ amounts
to twisted boundary conditions in the Hilbert space, while a normalized
sum over twisted boundary conditions in periodic Euclidean time can be
shown to be equivalent to a projector 
\BE						\label{GinvProj}
	\P_G={1\0|G|}\sum_{g\in G}\tbc g *	% \cdot	
\EE
onto $G$-invariant states. Under modular transformations
\BE	\textstyle
	\t\to{a\t+b\0c\t+d},~~~~
	\left({a~b\atop c~d}\right)~\in~\mao{PSL}(2,\IZ)
\EE
boundary conditions are recombined: For the generators
\BE
	S:\t\to -1/\t,\qquad~ T:\t\to \t+1
\EE
of $\mao{PSL}(2,\IZ)$ we observe
\BE
	S: \tbc g h ~\to~\tbc {h^{-1}} {g},\qquad 
	T: \tbc g h ~\to~\tbc {gh} h
\EE
where $T$ maps the double-periodicity $(1,\t)$ to $(1,\t+1)$ and the action of
$S$ has been chosen as $(1,\t)\to(\t,-1)$.%
\footnote{~
	While $S^2=(ST)^3=\id\in \mao{PSL}(2,\IZ)$ for modular group elements,
	the action of $S^2=(ST)^3:(1,\t)\to(-1,-\t)$ on the world sheet 
	amounts to parity plus time reversal. Due to CPT invariance the 
	action of $S$ on the Hilbert space thus squares to a charge 
	conjugation $S^2=(ST)^3=C$ of the conformal fields.
}
The double-periodicities are consistently defined only if $g$
and $h$ commute so that we need to restrict to twists obeying $gh=hg$ in
the case of non-abelian groups.

Since modular transformations % completely 
mix up all twists of the periodicities
along the homology cycles we expect an invariant to contain contributions
from all combinations and it is easy to see that the simplest invariant
solution is
\BE
	Z_{\cc/G}\equiv{1\0|G|}\sum_{gh=hg}\tbc gh 
%	\;=\; \sum_{C_i}{1\0|N_i|} \sum_{g\in N_i}~~\tbc {g~~}{~h\in C_i} 
\EE
In the abelian case the sum over $h$ corresponds to a sum over all twisted
sectors. The sum over $g$ then implements the projection onto invariant
states; in accord with (\ref{GinvProj}) the normalization ensures that the 
(invariant) ground state contributes to the partition function with 
multiplicity one.%
% the in accord with the projector (\ref{GinvProj}).%
\footnote{~
	With the restriction to $gh=hg$ the formula also applies to the
	non-abelian case, where the sum can be interpreted to extend over 
	conjugacy classes of twists followed by a projection onto states
	that are invariant under the respective normalizers.
}
%% \BE Z_{CFT/G}\equiv{1\0|G|}\sum_{gh=hg}\tbc gh \;=\; \sum_{C_i}{1\0|N_i|}
%%     \sum_{g\in N_i}~~\tbc {g~~}{~h\in C_i} 
%% \EE
%% stabilizer (little) group $N_i$ of $h$ in conjugacy class $C_i$ 
%% ($|N_i||C_i|=|G|$).\\
%% correlation fns.: L.Dixon, D.Friedan, E.Martinec, S.Shenker, NPB282(87)13;
%% S.Hamadi, C.Vafa, NPB279(87)465\\
%% Dixon,Harvey,Vafa,Witten NPB261(85)678; B274(86)285
%
Our CFT result thus coincides with what we expect for closed strings on 
orbifolds $X/G$. But there might be further solutions. 

\subsection{Discrete torsion and quantum symmetries}

Let us start with the more general ansatz
\BE
	Z^\e_{\cc/G}\equiv{1\0|G|}\sum_{gh=hg}~\e(g,h)\,\tbc gh 
\EE
with weight $\e(g,h)$ for the $(g,h)$--twisted contribution. This
modification can also be motivated from geometry and is
called ``discrete torsion'' \cite{va86} because it is related to phase
factors $\e(g,h)$ due to $B$-field flux with only ``discrete'' values
allowed by $G$-invariance (the field strength $H=dB$ of the 2-form $B$ 
determines the ``torsion'' of the corresponding sigma model). With an
analysis of the modular invariance and factorization constraints%
\footnote{~
	On a genus $n$ surface the partition function depends on $2n$
	twists along homology cycles, with a corresponding prefactor	
	$\e(g_1,g_2;\ldots;g_{2n-1},g_{2n})$ that has to factorize into
	$\e(g_1,g_2)\ldots\e(g_{2n-1},g_{2n})$. The only condition in the
	analysis that has to be used beyond the torus is a Dehn twist at 
	genus 2.
}
Vafa \cite{va86} has shown that 
\BE							\label{eg1g2}
	\e(g,g)=\e(g,h)\e(h,g)=1,~~~\e(g_1g_2,h)=\e(g_1,h)\e(g_2,h).
\EE
Mathematically discrete torsion corresponds to an element
of the group cohomology $H^2(G,U(1))$. For abelian groups 
$G=\IZ_{n_1}\ex\ldots\ex\IZ_{n_r}$ with generators $g_i$ the most general
solution is parametrized by an arbitrary choice of the phases 
$\e(g_i,g_j)$ for $i<j$ obeying $\e(g_i,g_j)^{\mao{gcd}(n_i,n_j)}=1$.

The ambiguity of the orbifold CFT that is due to discrete torsion is quite 
easy to understand in the operator picture 
because the group action is originally
defined only in the untwisted sector. For the twisted sectors we do know
the group action on (untwisted) operators but the action on the twisted
ground states (and on the corresponding twist fields) is a priory subject to a 
choice. We can thus think of $\e(g,h)$ as an extra phase of the group 
action of $g$ in the $h$-twisted sector.

While the symmetry of the original CFT is lost after orbifolding 
because of the projection to invariant states, a new symmetry
emerges due to selection rules for operator products of twist fields 
$\S_{h_1}(z)\S_{h_2}(w)$, to which we only expect contributions of 
fields twisted by $h_1h_2$. The corresponding symmetry of the orbifold has 
been called quantum symmetry \cite{va891}.
In the abelian case the quantum symmetry is dual of the original
symmetry. Modding out the quantum symmetry of a $\IZ_n$-orbifold 
just gives us back the original CFT \cite{odt}. If we mod out two 
commuting group actions $\bra g_1,g_2\ket$ in two steps then the freedom due 
to discrete torsion $\e(g_2,g_1)$ can be recovered by combining the group 
action $g_2$ of the second orbifolding with an approprite power of the 
quantum symmetry $q_1$ that emerges from the $g_1$-twist in the first
orbifold. These ideas can be used to extend the Green-Plesser mirror 
construction of Gepner models%
\footnote{~
	More generally, we can consider arbitrary N=2 SCFTs for which
	mirror symmetry, i.e. right-moving charge conjugation, is equivalent
	to an orbifold \cite{odt}. This is the case for the large class
	of Landau-Ginzburg models for which a transversal potential exists
	whose number of monomials is equal to the number of fields
	\cite{Kreuzer:1994uc}, as was discovered by Berglund and H\"ubsch
	\cite{Berglund:1991pp}.
}
to arbitrary orbifolds with discrete torsion \cite{odt,ade}.

\subsection{Simple currents}

Simple currents are, in a sense, generalized free fields in
rational conformal field theories. For free bosons there is a shift
symmetry. When it is used for orbifolding the twisted sectors 
correspond to winding states. For free fermions a $\IZ_2$ symmetry is 
provided by the fermion number. In this case the twisted sector is the 
Ramond sector, with a cut in the punctured complex plane, and the 
projection to invariant states is the GSO projection. Simple currents,
as we will see, also come with discrete symmetries. Accordingly, they
can be used to construct new conformal field theories, which turn out
to be given in terms of the original characters but with a certain type of 
non-diagonal modular invariants.

\del
CFT:: simple current techniques of Schellekens and Yankielowits; while
orbifolds (with and without discrete torsion) turned out to be
useful generalizations and, in particular, providing mirror constructions
in some cases. On the occation of a visit by J\"urgen Fuchs at CERN
we studied the relation between the spectra []. At the same time Beatrice and
Schellekens tried to classify the most general simple current modular
invariants (SCMI) and observed a unversality of the total numbers that was 
independent of the precise combination of automorphisms and extensions that 
come with a SCMI. When a count of the number of orbifolds with discrete
torsions reproduced these numbers for the abelian symmetries that are
implied by the simple currents it was natural to expect a close relation,
which then found to be comprised in a very simple formula for the
most general SCMI that can directly be interpreted in the orbifold languge.
\enddel

We consider a rational conformal field theory, i.e. a CFT with
left- and right-moving chiral algebras $\ca_L$ and $\ca_R$ such that 
the conformal fields are combined into a finite number of representations
$\Ph_{i\bar k}$ of $\ca_L\otimes \ca_R$, where $i$ labels the representation 
of $\ca_L$. The chiral algebras contain the Virasoro algebra and possibly 
more. We may use the highest weight state, or primary field, in a conformal
family as its representative. It is important, however, to keep in mind 
that the conformal weight $h_i$ is well-defined for a primary field, but only 
defined modulo 1 for the conformal family.

The fusion algebra $\Ph_i\ex\Ph_j=\cn_{ij}{}^k\Ph_k$ of a rational CFT is the
commutative associative algebra whose non-negative integral structure 
constants $\cn_{ij}{}^k$ encode the fusion rules, i.e. the information of
which representations of the chiral algebra appear in operator product
expansions $\Ph_i(z)\Ph_j(w)$.%
\footnote{~
	Multiplicities $\cn_{ij}{}^k>1$ indicate contributions from 
	descendents in OPEs beyond the coefficients that are implied
	by the Ward identities of the chiral algebra. 
}
A simple current $J$ of a conformal field theory is a primary field that has 
a unique fusion product with all other primary fields \cite{sc90r}, i.e.
\BE
	J\times \Ph_j=\Ph_{(Jj)},\qquad 
		\,j\,\to\,Jj\,\to\,J^2j\,\to\,J^3j\ldots,
\EE
where we use the notation $Jj$ for the label of the fusion product of $J$
and $\Ph_j$. A simple current thus decomposes the field content of the CFT 
into orbits, which have finite length for a rational theory.

Since the OPE $J(z) \Ph_j(w)$ contains only fields from a single conformal
family, whose conformal weights can only differ by integers,
all expansion coefficients $(z-w)^{h_{Jj}-h_J-h_j}$ 
have the same monodromy $e^{-2\p i Q_J(\Ph_j)}$ with
\BE
	Q_J(\Ph_j)\equiv h_J+h_j-h_{Jj}~~~\mod~~1
\EE
about the expansion point $w$. The monodromy of $J(z)$ for a big circle
about the positions of $\Ph_j(w_j)$ and $\Ph_k(w_k)$ is the product of the 
two respective monodromies. Thus the phase transformation 
$e^{-2\p i Q_J}$ is 
compatible with operator products and defines a symmetry of the CFT. 
Before we come to the resulting orbifold CFTs, which correspond to the
simple current modular invariants, we need to collect some basic 
definitions and facts for simple currents \cite{sc90,sc90r}.

The order $N_J$ of a simple current $J$ is the length of the orbit of the 
identity $J^{N_J}=\Id$.
Because of associativity and commutativity of the fusion product the simple 
currents of a CFT form an abelian group $\cc$, which is called the center. 
The definition of the monodromy charge implies 
$Q_{J\ex K}(\Ph)\equiv Q_J(K\Ph)-Q_J(K)+Q_K(\Ph)$ modulo 1, so that
\BE
	Q_{J\ex K}(\Ph)	\equiv Q_J(\Ph)+Q_K(\Ph), ~~~~~~
	Q_{J^n}(\Ph)\equiv nQ_J(\Ph).
\EE
$Q_J(\Ph)$ is hence a multiple of $1/N_J$. It can be shown that the
charge quantum of $Q_J$ is indeed $1/N_J$,
%, i.e. charges $k/N_J$ are present in the spectrum for at least one value 
% of $k$ with $\gcd(k,N_J)=1$,
so that a simple current $J$ always comes with a discrete $\IZ_{N_J}$ phase
symmetry of the CFT (not every cyclic symmetry is generated by a simple 
current, though). The symbol $\equiv$ henceforth denotes equality modulo 
integers.

For the orbifolding of a CFT we may choose to mod out some subgroup of its
full symmetry group. Similarly, we now choose some fixed subgroup $\cg$ 
% of order $\prod N_i$ 
of the center $\cc$ of a CFT 
that is generated by independent simple currents $J_i$ of orders $N_i$.
We use the notation $[\a]=\prod J_i^{\a_i}$ and $Q_i=Q_{J_i}$,
where $\a_i$ are integers that are defined modulo $N_i$. 
Then we can parametrize the conformal weights and monodromy charges of all
simple currents in $\cg$ in terms
of a matrix $R_{ij}$ \cite{ga91},					\VS-5
\BE						\label{h}		\VS-5
     R_{ij}= \frac{r_{ij}}{N_i}\equiv Q_i(J_j) =  Q_j(J_i), ~~~~
     h_{[\a]}\equiv \2\sum_i r_{ii}\a^i-\2\sum_{ij}\a^iR_{ij}\a^j\!
\EE
with $r_{ij}\in\IZ$.
If $N_i$ is odd we can always choose $r_{ii}$ to be even. With this convention
all diagonal elements $R_{ii}$ are defined modulo 2 for both, 
even and odd $N_i$.%
\footnote{~
	It is easiest to first compute $R_{ij}\equiv Q_i(J_j)$ modulo 1 and 
	then fix  $R_{ii}$ modulo 2 for the diagonal elements with even 
	$N_i$ by imposing that formula (\ref h) for $h(J_i)$ has to hold.
}
Using the definitions of $Q$ and $R$ we obtain				\VS-3
\BE						\label{Q}		\VS-3
     h_{[\a] \Ph}\equiv h_\Ph+h_{[\a]}-\a^iQ_i(\Ph), ~~~~~
     Q_i([\a] \Ph)\equiv Q_i(\Ph)+R_{ij}\a^j.
\EE
It can be shown that $S$ matrix elements for fields on the
same orbits are related by phases,				\VS-3
\BE								\VS-3
     S_{[\a]\Ph,[\b]\Ps}=
	S_{\Ph,\Ps}~e^{2\p i(\a^kQ_k(\Ps)+\b^kQ_k(\Ph)+\a^k R_{kl}\b^l)}.
\EE
$T$-matrix elements only depend on conformal weights and, according to
eq.\,(\ref Q), are related by phases $2\p i(h_{[\a]}-\a^iQ_i(\Ph)
	-h_{[\b]}+\b^iQ_i(\Ps))$.
%\BE
%	T_{[\a]\Ph,[\b]\Ps}=T_{\Ph,\Ps}~e^{2\p i(h_{[\a]}-\a^iQ_i(\Ph)
%	-h_{[\b]}+\b^iQ_i(\Ps))}
%\EE
%according to eq.\,(\ref Q) 
%along~orbits.

\subsection{Simple current modular invariants and chiral algebras} 

The partition function of a rational CFT can be written as 		\VS-3
\BE									\VS-3
	Z(\t)=\mao{Tr} e^{2\p i \t L_0}e^{-2\p i \bar\t \bar L_0}
	=\sum_{ij} M_{ij}\c_i(\t)\bar\c_j(\bar\t)
\EE
with a non-negative integral matrix $M_{ij}$ that is called a modular
invariant~if
\BE							\VS-3
		[M,S]=[M,T]=0\quad \hbox{and}\quad M_{\id\id}=1
\EE
since under modular transformations $\c_i(-1/\t)=S_{ij}\c_j(\t)$  
and $\c_i(\t+1)=T_{ij}\c_j(\t)$ so that $M\to S^tMS^*$ and $M\to T^tMT^*$
with symmetric unitary matrices $S$ and $T$, respectively.
Modular invariants of automorphism type are permutation matrices 
that uniquely map representation labels of the left movers to right movers,
where the permutation is an automorphism of the fusion rules.
Extension-type invariants, on the other hand, 
combine contributions of several characters to characters of 
extended chiral algebras while other representations of the original
chiral algebra %characters 
are projected out. % (see below). % may not contribute at all.

Simple current modular invariants (SCMIs) are modular invariants for which
$M_{jk}\neq0$ only if $\Ph_j$ and $\Ph_k$ are on the same orbit, i.e. 
if $k=Jj$ for some simple current $J\in\cc$. 
$T$-invariance requires that $h_j-h_k\in\IZ$, and is hence also called
``level matching''. Using eq. (\ref{Q}), with
the above notation $[\a]=\prod J_i^{\a_i}\in\cg\subseteq\cc$, we thus 
find the condition that						\VS-2
\BE								
	h_j-h_{[\a]j}\equiv\a^iQ_i(\Ph_j)-h_{[\a]}
	%=\a^iQ_i(\Ph_j)+\frac12\sum_{ij}\a^iR_{ij}\a^j-\frac12 r_{ii}\a^i
	~\in~\IZ	\label{level}
\EE
must be an integer.
If the order $N_i$ of $J_i$ is even then eq. (\ref{level}) implies that
the twist $J_i$ (like any odd power of $J_i$) can contribute to a modular 
invariant only if $r_{ii}=N_iR_{ii}\in2\IZ$. We henceforth assume that all
generators of $\cg$ satisfy this condition.%
\footnote{~
	The maximal subgroup of $%\cc'\subseteq
	\cc$ that can contribute to a SCMI is called ``effective center''.
}

If we	%% We can 
think of $[\a]$ as % corresponding to
the twist in the orbifolding procedure, which is in accord with the number
$|\cg|$ of twisted sectors as well as with the expected quantum symmetry
due to twist selection rules,
% With the orbifolding procedure in mind 
it is %now 
not difficult to guess
that the SCMI should impose a projection $\d_\IZ(Q_i+X_{ij}\a^j)$ where  
$\d_\IZ$ is one for integers and zero otherwise. The linear ansatz 
$X_{ij}\a^j$ for the phase shift in the projection is suggested by
comparing eq. (\ref{level}) with $h_{[\a]}\equiv-\frac12\a^iR_{ij}\a^j$ 
and % as well as 
by the expected quantum symmetry.
Using regularity assumptions%
\footnote{ ~
	`Regularity' requires that $M_{\Ph,[\a]\Ph}$ only depends on 
	$Q_i(\Ph)$
	\cite{ga91}. Discrete Fourier sum and 2-loop modular invariance 
	imply
	that the `phases' are bilinear and antisymmetric \cite{kr94}.
%	For an example where this condition is violated and other simple
%	current modular invariants exist
 %	(which, however, are not physically sensible) see \cite{fu94}.
}
it can be shown  \cite{kr94} that the most general SCMI reads		\VS-9
\BE \label{X}								\VS-12
     M_{\Ph,[\a]\Ph}=\m(\Ph)\prod_i\d_\IZ\left(Q_i(\Ph)+ X_{ij}\a^j\right),
\EE
where $T$-invariance implies $X+X^T\equiv R$ modulo 1 for off-diagoal and 
modulo 2 for diagonal matrix elements, $X$ is quantized by
$\gcd(N_i,N_j)X_{ij}\in\IZ$, and $\m(\Ph)$ denotes the multiplicity of the 
primary field $\Ph$ on its orbit, i.e. 
$\m(\Ph)=|\cg|/|\cg_\Ph|$ where $|\cg_\Ph|$ is the size of the orbit of the
action of $\cg$ on $\Ph$. While the symmetric part 
$X_{(ij)}\equiv\frac12R_{ij}$ of $X$ is fixed by level matching, the 
ambiguity due to the choice of a properly quantized antisymmetric part 
$E_{ij}\equiv X_{ij}-\frac12R_{ij}$ corresponds to the discrete torsion of the
orbifolding procedure.

We can now briefly discuss different types of invariants. If $X=0$ we have
a pure extension invariant because all fields with non-integral charges 
are projected out while all fields on a simple current orbit are combined
to new conformal families. $X=0$ is only possible if the conformal weights 
of all simple currents $J\in\cg$ are integral and since these currents are
in the orbit of the identity they extend the chiral algebras $\ca_L$ and
$\ca_R$ so that we obtain a new rational symmetric and diagonal CFT.

Let us define the kernel $\mao{Ker}_\IZ X$ as the set of integral solutions
$[\a]$ of $X_{ij}\a^j\in\IZ$ with $\a_j$ definded modulo $N_j$. If this kernel
is trivial then $\left(Q_i(\Ph)+ X_{ij}\a^j\right)\in\IZ$ has a unique
solution $[\a]$ for each charge, which defines a unique position $[\a]\Ph$
on the orbit that only depends on the charge $Q_i(\Ph)$ of $\Ph$. We then
obtain an automorphism invariant. In general, the extension of the
right-moving chiral algebra $\ca_R$ is give by the kernel $\mao{Ker}_\IZ X$
and, since					\VS-5
\BE \label{XT}						\VS-4
     M_{[\a]\Ph,\Ph}=%\mao{Mult}
	\m(\Ph)\prod_i\d_\IZ\left(Q_i(\Ph)+\a^jX_{ji}\right),
\EE
the extension of the left-moving chiral algebra $\ca_L$ is give by the 
kernel $\mao{Ker}_\IZ X^T$ of the transposed matrix. While the extensions
are of the same size, they need not be isomorphic. %
%\footnote{~
	For example, an extension of $\ca_R$ by $\IZ_9$ can occur together
	with an extension of $\ca_L$ by $\IZ_3\ex\IZ_3$.
%}

\section{Gepner-type (0,2) models}	\label{sec:3}

The right-moving sector of a heterotic string consists of four space-time
coordates and their superpartners $(X^\m,\ps^\m)$, a ghost plus superghost 
system $(b,c,\b,\g)$, and a supersymmetric sigma model on a Calabi-Yau, 
whose abstract version is an $N=2,c=9$ SCFT $\cc_{int}$. Equivalently, we 
can use light-cone gauge, which amounts to ignoring the ghosts and 
restricting space-time indices to transverse directions.
The left-moving sector is a bosonic string with space-time plus ghost part
$(X^\m,b,c)$ and the same internal sector $\cc_{int}$ with $c=9$, whose
central charges add up to $4+9-26=-13$ so that we need to add a left-moving
CFT with central charge 13 for criticality. Modular invariance requires
this CFT to be either an $E_8\times SO(10)$ or $SO(26)$ level 1
affine Lie algebra (we will henceforth ignore the $SO(26)$ case because it
is phenomenologically less attractive). In the geometric context of a 
sigma model on a Calabi-Yau the superstring vacuum is then 
obtained by aligning space-time spinors and tensors with 
internal Ramond  and Neveu-Schwarz sectors, respectively,
and performing the GSO projection. For abstract $N=2$ SCFTs U(1) charges
may be quantized in fractional units so that, in addition, a projection to
integral charges (generalized GSO) is required for space-time supersymmetry. 

All of these operations can be understood as SCMIs of extension type
\cite{sc90,kr95}.
To see this let us first discuss the simple currents in the relevant 
CFTs. For the $D_n\cong SO(2n)$ current algebra the center $\cc_n$ has 
order 4 and consists of the spinor representation $s$, its conjugate $\sc$, 
and the vector $v$ with				\VS-7
\BE						\VS-4
	sv=\sc,~~~s^2=\sc^2=v^n,~~~v^2=\id~~\then~~\cc_n\cong
	\hbox{\fns$\begin{cases}~~~\IZ_4&\hbox{ for }n\not\in2\IZ %D_{2l-1}
			\\\IZ_2\ex\IZ_2&\hbox{ for }n\in2\IZ %D_{2l}
		\end{cases}$}.
\EE
%with 
The conformal weights and monodromies are
\BE	\textstyle
	h_s=\frac n8,~~h_v=\2,~~~%N_s=\{{2~~(n{\rm~even})\atop4~~(n{\rm~odd})}
	R_{vv}=1,~~R_{vs}=1/2,~~R_{ss}=\hbox{\fns$
		\begin{cases}\!\,3n/4&\hbox{ for }n\not\in2\IZ %\hbox{ odd}
				\\~n/4&\hbox{ for }n\in2\IZ %\hbox{ even}
		\end{cases}$}
	% =\{{n/4~~(n{\rm~even})\atop3n/4~~(n{\rm~odd})}
\EE
since $s^2=v^n$ so that $N_s=4$ for $n$ odd and $N_s=2$ for $n$ even.

For the internal $N=2$ SCFT 
									\del
the most important states are the 
(anti)chiral primary states, which satisfy the BPS condition $h=\pm\2Q$,
and the Ramond ground states with $h=c/24$. They saturate the unitarity
bounds\cite{lvw} that follow from the $N=2$ algebra
\BEA
&&\HS-12	\hbox{\small
	$\{G_r^-,G_s^+\}=2L_{r+s}-(r-s)J_{r+s}+\8{c\03}(r^2-\8{1\04})\d_{r+s},
      	~~~~ [L_n,G_r^\pm]=(\8{n\02}-r)G_{n+r}^\pm$},			\nn\\
&&\HS-12	\hbox{\small$
	[J_m,J_n]=\8{c\03}m\d_{m+n},~~~~ [J_n,G_r^\pm]=\pm G_{n+r}^\pm,
	~~~~ [L_n,J_m]=-mJ_{m+n} $},
\EEA
This algebra admits the continous spectral flow
\BE \textstyle\hbox{\small$
	L_n\rel {\cu_\th} \longrightarrow	%\cu_\th^{-1}L_n\cu_\th=
	L_n+\th J_n+{c\06}\th^2\d_n,~~~~~
    J_n\rel {\cu_\th} \longrightarrow	%\cu_\th^{-1}J_n\cu_\th=
	J_n+{c\03}\th\d_n,~~~~~
	G_r\rel {\cu_\th} \longrightarrow%    \cu_\th^{-1}G_r^\pm\cu_\th=
		G_{r\pm\th}^\pm$} 
\EE
which for $\th=\pm\2$ maps Ramond ground states into chiral and antichiral
primary fields, respectively.
								\enddel
$\cc_{int}$ the center always contains the supercurrent $J_\su$ with
$h=3/2$ and $J_\su^2=\id$ and the spectral flow current $J_\sp$ with
$h=c/24$ and $J_\sp^{2M}=J_\su^k$ where $c=3k/M$ and $1/M$ is the charge
quantum in the NS sector (see appendix B) \cite{kr95}. The monodromy charge
$Q_\su$ is 0 in the NS sector and $1/2$ in the Ramond sector. 
$J_\sp=e^{i\sqrt{c/12}X}$ is the Ramond ground state of maximal $U(1)$
charge $c/6$ and can be written as a vertex operator in terms of the 
bosonized $U(1)$ current \cite{lvw} $J(z)=\sqrt{c/3}\,\6X(z)$ so that 
$Q_\Jsp\equiv-\frac12Q$ and $Q_\Jsp(J_\sp)\equiv-c/12$ modulo 1.

% we observe that $\co_q=e^{i\sqrt{3/c}qX}\co_0(\6\Ph,\ps_0,\ldots)$\\
%$\cu_\th(z)\Ph_q(z',\5z')\sim(z-z')^{q\th}\Ph_{q+c\th/3}$;
%contribution of $q$ to $h$ is $3q^2\02c$.\\

\del
%\begin{figure}
\hrule\hrule \VS-9
{
\footnotesize
\noindent
\begin{tabbing}
Simple currents of $D_n$:~	
	\=$J_v^2=\Id$,~ $J_sJ_v=J_c$,
	\=~~~$h_v=1/2$,~~~~
	\=$N_v=2$,~~~~~
	\=~~~$R_{vv}=1$~~~$R_{vs}=1/2$
\\[4pt]
~~~$\mao{Center}(D_{2n})\simeq\IZ_2\ex\IZ_2\atop
		\mao{Center}(D_{2n-1})\simeq\IZ_4\HS2 $	
	\>$J_s^2=J_c^2=J_v^n$,
	\>\HS-12 $h_s=n/8=h_c$,
	\>$N_s=\{{2~~(n{\rm~even})\atop4~~(n{\rm~odd})}$
	\>\HS33 $R_{ss}=\{{n/4~~(n{\rm~even})\atop3n/4~~(n{\rm~odd})}$
\end{tabbing}}
\enddel

\subsection{The (2,2) case and the generalized GSO projection}

In order to apply simple current techniques it is convenient to start
from a left-right symmetric theory. This can be achived by applying the
bosonic string map to the right-movers \cite{sc90},		\VS-3
\BE								\VS-3
	SO(2)_{LC} \to D_5\ex E_8,\qquad
	(0,v)\to(v,0), \quad (s,\sc)\to-(\sc,s),	\label{BSM}
\EE
which maps modular invariant partition functions of heterotic strings to 
modular invariant partition functions of bosonic strings. The inverse
map will be called Gepner map.
For simplicity we discuss the spectrum in terms of light-cone space-time 
$SO(2)_{LC}$ representations rather than using the equivalent 
$SO(4)\otimes(b,c,\b,\g)$, which would necessitate superghosts 
contributions with the benefit of manifest Lorentz invariance.

Consistent quantization of the gauge fixed N=1 supergravity theory requires
that the Ramond and NS sectors of the space-time and internal sectors are
aligned. After the bosonic string map this implies that $SO(10)$ spinor
representations are aligned with the Ramond sector of the internal %$c=9~N=2$
SCFT. This can be implemented by a SCMI that extends the chiral
algebra by the current $J_{RNS}=J_v\otimes v$ (which has conformal weight 
$h_{RNS}=2$) because $Q_{J_v}\equiv 1/2$ for Ramond fields and 
$Q_v\equiv 1/2$ for $SO(10)$ spinors. Similarly, in the case of a Gepner 
model, where 
$\cc_{int}=\cc_{k_1}\otimes\ldots\otimes \cc_{k_l}$ is a tensor product of
$N=2$ SCFTs, the alignment can be implemented as a SCMI extending the 
chiral algebra by all bilinears of the respective supercurrents 
$J_{ij}=J_{v_i}J_{v_j}$, where $h_{ij}=3$. Rather then defining a 
``superconformal tensor product'' with an implicit alignment we keep the 
alignment procedure explicit because we will later be interested in (0,2) 
models for which the chiral algebra extension that implements the alignment 
only takes place in the right-moving sector, where it is needed for
consistency.

Space-time supersymmetry now requires that the spectral flow in the internal
sector is combined with an $SO(10)$ spin field $s$ after the bosonic string
map so that space-time bosons/fermions in the heterotic string
have NS/R contributions from the internal N=2 SCFT \cite{ba88}.
This is implemented by the simple current $J_{GSO}=J_\sp\otimes s$, which 
has integral conformal weight $h_{GSO}=c/24+n/8=3/8+5/8=1$ and hence
can be used for a SCMI of extension type. Inspection of the massless
spectrum (see below) shows that the $2\times 16$ states in
$(J_{GSO})^{\pm1}$ together with the $U(1)$ current of the $N=2$ SCFT lead
to the 33 massless vector bosons that extend the $45_{adj}$ of $D_5$ to
the $78_{adj}$ of the gauge group $E_6$ that is familiar from 
the standard embedding $SU(3)\subseteq E_8$.
The mechanism that implements space-time SUSY in the fermionic string is 
hence related by the bosonic string map to the mechanism that extends 
$E_8\times SO(10)$ to the gauge group $E_8\times E_6$ of a (2,2) 
compactification. Since $Q_{GSO}=-\frac 12 Q$ this 
``generalized GSO projection''
implies a projection to even $U(1)$ charges in the bosonic string and,
according to eq. (\ref{BSM}), to odd $U(1)$ charges in the Gepner construction
of the superstring \cite{Gepner} when the space-time contribution is taken 
into account.

For sigma models on CY manifolds the charges are already quantized in
(half)integral units in the (R)NS sector. The standard GSO projection 
can hence be regarded as a generalized GSO projection with $M=1$. 
In order to simplify the comparison between abstract and geometrical constructions 
of $N=2$ SCFTs it has been suggested to define an
intermediate projection which extends the chiral algebra only by simple
currents that have no contributions from the spacetime/gauge sector 
\cite{Fuchs:2000gv}.
The corresponding subgroup $\cg_{CY}$ of the center contains all alignment
currents of the building blocks of the internal SCFT plus the current
$J_{CY}=J_{GSO}^2J_{RNS}^{c/3}=J_s^2J_v^{c/3}$.%
\footnote{~	The discussion in ref. \cite{Fuchs:2000gv} attemts
	independence of the space-time dimension $2n=10-2c/3$. Note, however, 
	that standard compactifications on K3's
	have internal $N=4$ SCFTs so that the bosonic analog of 
	$\cn\!\!=\!\!2$ 
	space-time SUSY in 6-dimensional (4,4) models 
	is the extension of the gauge group $E_8\times D_6$ to 
	$E_8\times E_7$, where the $3=133-66-2\cdot 32$ $D_6$-singlet gauge
	bosons come from the $SU(2)$ R-symmetry currents of the $N=4$ SCFT.
}

\del
For Calabi--Yau (CY) compactifications the string vacuum is obtained from 
the tensor product of the space-time with the internal sector by aligning
spin structures (i.e. the respective Ramond and Neveu-Schwarz sectors) 
and by the GSO projection. The spin structure alignment is necessary for
world-sheet supersymmetry (i.e. a consistent decoupling of ghosts via BRST)
and can be implemented as a simple current extension by the alignment current
$J_A=J_\su^{int}\otimes v$ where $v$ is the vector of $D_n$ (with $n=5$
after the bosonic string map). As expected, $h_A=2$ is integra, and the 
monodromy charge $Q_A$ is indeed integral for internal Ramond states times
space-time spinors and for internal NS states times space-time tensors. 
\enddel

In order to set up the enumeration of massless states of the heterotic 
string we recall the relevant vertex operators. On the bosonic side, 
where the NS vacuum has $h=-1$, there are the universal operators 	\VS-5
\BE								\VS-5
	\left(\6X^\m\ex \id_{E_8\ex D_5}+\id_{st}\ex 
	J_{-1}^{(E_8\ex D_5)}\right)\ex\id_{int}
\EE
and the model-dependent contributions				\VS-5
\BE								\VS-5
	 \id_{st}~\ex \id_{E_8}~\ex \!\!\!\!\!\!
	\sum\limits_{0\le r <4 \atop h_{int}=1-h_{D_5}(s^r)} \!\!\!\!\!
         (s)^r\ex \Ph_{int}		\label{model-dependent}	
\EE
For the right-movers
the NS vacuum has $h=-1/2$ and the relevant 
vertex operators are						\VS-11
\BE									\VS-1
	\sum\limits_{0\le r <4 \atop \5h_{int}=1/2-h_{D_1}(s^r)} \!\!\!\!\!
	\5{(s)}\,^r_{st}\ex \5\Ph_{int}.
\EE
The enumeration of the non-universal states can therefore be organized
according to the following data,
\begin{center}\fns\VS3
\begin{tabular}{||c|cc\TVR{4.2}{1.8}||} \hline\hline
	$D_5^{(B)}$ & $h_{int}$ & $Q_{int}$ \\\hline  
	$0=~\id\,$ & $1$ & $\pm2,0$\\  $s=\9{16}$ & $3\08$&$\!{3\02},-\2$ \\ 
	$v=\9{10}$&$\2$ &$\pm1$\\\VR03$\sc=\5{\9{16}}$&$3\08$&$\!\2,-{3\02}$\\
\hline\hline\end{tabular}
\HS8
	{\fns\def\H{@{\hspace{3.5pt}\VR{2}0}}%
\begin{tabular}{c\H c\H c\H c\H c\H c\H c}	\unitlength=2.8pt 
	\BP(0,0)(2,0) 
	\putvec(7,-20,-1,1,6)	\putvec(18.5,-20,1,1,6)	
	\putlab(2,-18)t{$Q$}	\putlab(24,-18)t{$\5Q$}
	\EP
	&&& 1 &&&\\ && $\,y\,$ && $\,x\,$ &&\\ & $\,y\,$ && $a$ && $\,x\,$ &\\
	1~&& $g$ && $g$ &&1\\
	& $\,x\,$ && $a$ && $\,y\,$ &\\ && $\,x\,$ && $\,y\,$ &&\\ &&& 1 &&&\\
\end{tabular}}%
\HS8
\begin{tabular}{||c|cc\TVR{4.2}{1.8}||} \hline\hline
	$\!\!D_5$ $\rightarrow$ $D_1^{(F)}$ & $\5h_{int}$ &$\5Q_{int}$\\\hline
	$0$ $\rightarrow$ $\!{\Ps^{\m}}\!=v$  & $0$ & $0$\\   
	$s$ $\rightarrow$ $\:\9{\5\S}=\sc\;$  & $3\08$ & $\!\!{3\02},-\2$\\
	$v$ $\rightarrow$ $\:\id\!_{_{st}}\!=0$  & $\2$ & $\pm1$\\  
	$\sc$ $\rightarrow$ $\:\9\S=s\;$  & $3\08$&$\!\!\2,-{3\02}$\VR03\\
\hline\hline
\end{tabular}
\end{center}\VS3
where the entries of the ``Hodge diamond'' are multiplicities of 
internal fields with (left,right) charges $(Q,\bar Q)$.

Since spectral flow relates (anti)chiral primary states to Ramond ground 
states the counting can be performed in any of these sectors, with an
appropriate shift of charges. 
% Charge conjugation in the Ramond sector of $c=9$ $N=2$ SCFTs constrains 
% the entries as indicated. 
For CY compactifictions Hodge duality further implies $x=y$ where $y=1$
corresponds to extended $\cn=2$ space-time SUSY and $y=3$ yields $\cn=4$.
The bosonic (left-moving) analogs of these extensions are gauge groups 
$E_7$ and $E_8$, respectively.
For orbifolds with discrete torsion $x\neq y$, i.e. any combination of
$E_{6,7,8}$ with $\cn=1,2,4$, is possible \cite{odt,ade,lgt}.
%, where $x\neq y$ is possible for orbifolds with 
%discrete torsion \cite{odt,ade,lgt}.
The $h_{12}=a$ complex structure deformations (we call them % $a$ 
{\it anti-generations} of charged particles) 
correspond to chiral primary fields with symmetric charges $Q=\bar Q=1$
while the $h_{11}=g$ 
{\it generations}  count K\"ahler moduli, i.e. the CY Hodge
diamond is rotated by $\p/2$ as compared to the diamond of left/right charge 
multiplicities of the $N=2$ SCFT.

\subsection{The extended Poincar\'e polynomial}
 
The aim of the extended Poincar\'e polynomial (EPP) is to encode all 
information about an $N=2$ superconformal theory that is necessary for 
computing the (charged) massless spectrum of any tensor product	  % with $c=9$
containing this model as one factor. 
It takes advantage of the fact that the generalized GSO-projection
corresponds to an extension invariant so that we may, in a first step, 
disregard the projection to integral charge in the expression (\ref X)
and consider the `unprojected orbifold'. Eventually, to obtain the 
projected orbifold, we just have to omit the 
contributions with non-integral monodromy charges.

The Poincar\'e polynomial encodes charge degeneracies for % For 
$N=2$ SCFTs, %% the Poincar\'e polynomial encodes %the charge degeneracies,
\BE
	P(t,\5t)=\tr_{(c,c)} t^Q\,\5t{}^{\5Q}
	=(t\5t)^{c/6}\tr_{R_{gs}} t^Q\,\5t{}^{\5Q},
\EE
where we assume locality of symmetric spectral flow.
In order to be able to combine the information of the factors of a tensor 
product we need to encode, in addition, information on the twists. 
We thus define the `full extended' Poincar\'e polynomial as		\VS-4
\BE 							\label{fepp}	\VS-4
     \cp(t,\5t,x,\s)=\sum_{l\ge0}~\sum_{k=0}^1x^l\s^kP_{l,k}(t,\5t),
\EE
where $P_{l,k}(t,\5t)$ is the Poincar\'e polynomial 
of the unprojected sector twisted by
$J_s^{2l}J_v^k$, i.e. $P_{l,k}$ is obtained by looking for all 
Ramond ground states $\Ph_{ij}$ with $j=J_s^{2l}J_v^ki$ and the $U(1)$ 
charges of $i$ and $j$ are encoded by the exponents of $t$ and $\5t$, 
respectively.

\del
The information on the location of Ramond ground states 
on the simple current orbits of 
$J_s$ and $J_v$ is important if
we consider tensor products of $N=2$ factor theories. For simplicity we 
restrict ourselves to the case that the tensor product contains
only two factors; what we are really interested in in this situation
is the modular invariant obtained for the total spinor
current $J_s^{(1)}J_s^{(2)}$ after the alignment of R and NS sectors. 
\enddel
For a 
tensor product with alignement of Ramond/NS sectors we obtain 
% the `full extended' Poincar\'e polynomial by the prescription
\[	% \textstyle
     \cp(t,\5t,x,\s)=\sum\limits_{l\ge0} x^l \Bigl(
	\sum\limits_{k=0}^1P^{(1)}_{l,k}(t,\5t)P^{(2)}_{l,k}(t,\5t)+
     \s\!\sum\limits_{k=0}^1P^{(1)}_{l,k}(t,\5t)P^{(2)}_{l,1-k}(t,\5t)\Bigr)
\]
By iteration of this formula we conclude that (\ref{fepp})
indeed encodes all information from the
factor theories of a Gepner model that enters the computation of the 
charged massless spectrum. In fact, this information is still redundant: 
Consider a R ground state $\Ph_{ij}$ whose contribution to $P_{l,k}$ is 
$t^{Q+{c\06}}\,\5t{}^{\5Q+{c\06}}$. 
Then eqs.\ (\ref Q) and (\ref{MMMM}) imply for the $U(1)$ charges
\BE 
     \5Q\eq Q+l\,c/3-k\quad\then\quad k\eq Q+l\,{c/3}-\5Q ~~~\mod ~ 2.
				\label{sign} 
\EE
As the exponent of $\s$ is fixed in terms of the other exponents
we can set 
\BE
	\s\to-1\quad\then\quad	\cp(t,\5t,x):=\cp(t,\5t,x,-1).
\EE
The negative sign is convenient for index computations since it implies
opposite signs for contributions to generations and anti-generations.%
\footnote{~
	In the original definition of the extended Poincar\'e polynomial 
	\cite{sc91} Schellekens, in addition,
	puts $\5t=1$. For diagonal theories we have shown \cite{kr95}
	that, for a given $Q$, all states contribute with the same sign, so 
	that it is indeed sufficient to drop the $\5Q$-dependence in 
	applications to heterotic (2,2) string vacua built from diagonal 
	theories, but not necessarily for orbifolds thereof.
}
For minimal models at level $k=K-2$ one finds  \cite{kr95}	\VS-4
\BE	\textstyle						\VS-4
	\cp^{(MM)}(t^K\!\!,\,\5t^K\!\!,x)=\sum\limits_{l=1}^{K-1}(t\5t)^{l-1}
		{1-(-x)^l~\5t^{K-2l}	\0	1-(-x)^K}
	=\frac{P(t\5t) 
	- \sum\limits_{l=1}^{K-1}(-x)^lt^{l-1}\5t^{K-1-l}}{1-(-x)^K}
\EE
where the ordinary Poincar\'e polynomial is $P(t^K)={1-t^{K-1}\01-t}$. 

Since the numbers of (anti)chiral primaries and of Ramond ground 
states are finite also in non-rational SCFTs extended
Poincar\'e polynomials can be defined in a more general context and
explicit formulas have been given for Landau-Ginzburg orbifolds \cite{kr95}.

\subsection{Gauge/SUSY breaking and (0,2) models}

While the chiral algebra extension of a SCMI based on $J_{GSO}$ and 
alignment currents %$J_{RNS}$ and $J_{ij}$ 
can be reduced by switching on discrete torsion $X\neq X^T$ this would
not only break the left-moving $E_6$ but also the right-moving space-time
SUSY of the heterotic string. We hence need to increase the twist group 
$\cg$ at least by one additional generator of even order. While there
are many possibilities for this type of models we would always end up 
with at least $SO(10)$. For smaller gauge groups, like the ``exceptional''
series $E_5=D_5=SO(10)$, $E_4=A_4=SU(5)$ and $E_3=SU(3)\times SU(2)$ that
is familiar from geometric/sigma model constructions, we have to start with
smaller building blocks and use asymmetric extensions
that rebuild the $D_5\times E_8$ needed for the Gepner map only in
the right-moving sector.

A natural implementation of this idea can be motivated by the free fermion 
construction of $D_n=SO(2n)$ in terms of $2n$ Majorana fermions with 
aligned spin structures.
The extension of $SO(2m)\otimes SO(2n)$ to $SO(2m+2n)$ is %then 
achived 
by aligment of all spin structures and can be implemented by a SCMI of 
extension type with the current $J\!=\!v_{_{D_m}}\!\otimes\! v_{_{D_n}}$,
in complete analogy to the alignment of spin structures for a tensor product
of SCFTs. The exceptional series is thus obtained by starting with
a gauge sector $SO(2l)\otimes SO(2)^{5-l}\otimes E_8$ and a generalized
GSO projection  \cite{bl95}
\BE
	J_{GSO}=J_s \otimes s_{_{SO(2l)}}\otimes (s_{_{SO(2)}})^{5-l}
\EE
as is illustrated in the following table:
\begin{table}[h]
\BC	\scriptsize	\VS-2
\begin{tabular}{||\TVR{3.5}{1.4}c|l|c|c|l||}			\hline$
l $&$ E_{l+1} $ & $ D_{l}\ex D_1^{5-l} $ &  $ 
	|E_{l+1}|-|D_{l}|-|U(1)|	$ & $
{\rm currens~} (J_{GSO})^{\pm1}			$\\\hline\hline$
5 $&$ E_6	 $&$ SO_{10} $&$ 32=78-45-1	$&$ |s|=16 %
		~~~ h={5\08}+{3\08}		$\\\hline$
4 $&$ E_5=SO_{10}     $&$ SO_8\ex SO_2  $&$ 16=45-28-1	$&$ 
	|s|=8	~~~ h={4\08}+{1+3\08}	$\\\hline$		%% 
3 $&$ E_4=SU_5	      $&$ SO_6\ex (SO_2)^2  $&$ ~8=24-15-1	$&$ 
	|s|=4	~~~ h={3\08}+{2\ex1+3\08}$\\\hline$		%%
2 $&$ SU_3\ex SU_2$&$ SO_4\ex (SO_2)^3  $&$ ~4=11-6-1~	$&$ 
	|s|=2	~~~ h={2\08}+{3\ex1+3\08}$\\\hline		%%
\end{tabular}	\VS-22	
\EC
\end{table}
\del
We know that there are no other vectors coming from the $J_s^{tot}$ orbits
since $J_s^{\pm1}$ are the only intersections of this orbit with the R
ground states for minimal models. In the NS sector the weight of the
contributions from the gauge sector is probably too large. 

\subsection{(0,2) vacua and generalization of the Bonn model}	

To make possible the bosonic string map we extend $SO_{16-2r}\ex (SO_2)^{r-3}$
to $SO(10)$ by $r-3$ simple currents $J_{ext}=J_v^{D_1}\ex J_v^{D_{8-r}}$, 
i.e. we put all bilinears in the vector currents into the anti-chiral algebra.
In addition we need $J_s^{tot}=J_s^{D_{8-r}}\ex \prod_1^{r-3}J_s^{D_1}\ex 
J_s^{c=9}$ in the anti-chiral algebra (GSO projection).

For phenomenological reasons we want to extend the gauge group
$SO_{16-2r}\ex(SO_2)^{r-3}\ex U(1)^{c=9}$ to $E_{9-r}$, so we
want $J_s^{tot}$ also in the chiral algebra.
$J_s^{tot}$ %=J_s^{D_{8-r}}\ex \prod_1^{r-3}J_s^{D_1}\ex J_s^{c=9}$ which 
has $h={8-r\08}+{r-3\08}+{9\024}=1$ and $N=4M$, where $M$ is the inverse 
charge quantum in the NS sector of the internal $c=9$ $N=2$ superconformal 
field theory. Of course, in addition to $E_{9-r}$ we will have $r-3$ extra
orthogonal $U_1$'s. 
\enddel

\noindent
For the rest of this paper we restrict to the case $l=4$, i.e. to $SO(10)$
models based on a CFT of the form
$\cc_{int}\times SO(8)\times SO(2)\times E_8$ with $c=26-4$.

Blumenhagen and A. Wi{\ss}kirchen
\cite{bl95} performed a computer search for spectra of heterotic models 
of this type that agree with Distler-Kachru models and came up with a 
small list, the most promising candidate of which is an SO(10) model with 
80 generations.  
They used the original approach of Schellekens and Yankielowicz
constructing SCMIs as products of invariants for cyclic subgroups of the 
center \cite{sc90}.
% They used the stochastic approach of Schellekens and Yankielowicz \cite{sc90}
% with the original (redundant) representation of the modular invariant as a 
% product of invariants for cyclic subgroups of the center. 
Translating their data into our language
we find, in addition to $J_{GSO}$ and the alignment currents,
a $\IZ_4$ twist whose simple current generator
$J_B=(J_s^{k=3})^5\ex s_{_{SO(2)}}$ is the
product of the spinor of $SO(2)$ times the 5$^{th}$ power of the spectral
flow of one of the minimal model factors of the quintic.

We call $J_B$, which squares to the alignment current 
$J_B^2\!=\!J_v^{k=3}\otimes v_{_{SO(2)}}$, {\it Bonn twist}. Since only 
one minimal model enters this construction it appears natural
to generalize the discussion to an internal SCFT of the form \cite{kr96} 
$\cc_{int}=\cc'\otimes \cf_K$, where $\cf_K$ is a minimal model whose level 
$k=K-2$ needs to be odd in order that $J_s^{2K}=J_v$. In the Landau-Ginzburg 
discription $\cf_K$ has a Fermat-type potential $W=\Ph^K$ and is hence 
referred to as {\it Fermat factor}. The Bonn twist thus generalizes to
\BE
	J_B=(J_s^{\cf})^K\ex s_{_{SO(2)}}, \qquad N_B=4,
			\quad J_B^2=J_v^{\cf}\otimes v_{_{SO(2)}}
\EE
so that the resulting (0,2) model can be defined by a SCMI based on the
generators $J_B$, $J_{GSO}$ and two more alignment currents
\BE
	J_A=v_{_{SO(8)}}\otimes v_{_{SO(2)}},\qquad
	J_C=J_v^{\cc'}\otimes v_{_{SO(8)}}.
\EE
The nonvanishing monodromies are $R_{BB}\equiv{K-1\02}\mod2$, 
$R_{AB}%=R_{BA}
\equiv\2\mod1$ and $R_{B,GSO}\equiv{K-1\04}\mod1$. 
We need $J_\GSO$ and the alignment currents $J_A$, $J_B^2$ and $J_\X$ in the
chiral algebra on the right-moving side, i.e. in the kernel of $X$, so that
the corresponding columns of the matrix $X$ must be 0 $\mod 1$, or 0
$\mod 1/2$ in the case of $J_B$. 		% $1/2$). 
% This fixes all freedom in the discrete torsions as shown in table 2.
%The (2,2) Gepner model corresponds to the twist group generated by
% $J_{GSO}$, $J_A$, $(J_B)^2$ and $J_C$ so that these currents need to
% enter the extended chiral algebra $\ca_R$, i.e. in the kernel of $X$, 
% so that the corresponding rows of the matrix $X$ must be 0
% (in the case of $J_B$ modulo $1/2$).
This fixes all discrete torsions % $X-X^T$ 
and implies
\begin{center}\scriptsize
\begin{tabular}{||c||c|c|c|c\TVR4{1.5}||} \hline\hline
$R$     & $J_\GSO$& $J_A$ & $J_B$       & $J_\X$ \\ \hline\hline
$J_\GSO$        & 0     & 0     & $K-1\04$ & 0 \\ \hline
$J_A$   & 0     & 0     & $\2$  & 0 \\ \hline
$J_B$   & $K-1\04$ & $\2$ & $K-1\02$    & 0 \\ \hline
$J_\X$  & 0     & 0     & 0     & 0 \\ \hline\hline
\end{tabular}
~~~~
\begin{tabular}{||c||c|c|c|c\TVR4{1.5}||} \hline\hline
$X$     & $J_\GSO$& $J_A$ & $J_B$       & $J_\X$ \\ \hline\hline
$J_\GSO$        & 0     & 0     & $K-1\04$ & 0 \\ \hline
$J_A$   & 0     & 0     & $\2$  & 0 \\ \hline
$J_B$   & 0     & 0     & $K-1\04$ & 0 \\ \hline
$J_\X$  & 0     & 0     & 0     & 0 \\ \hline\hline
\end{tabular}\ns
%\\[12pt]{\bf Table 2:} Monodromies and torsions for the modular invariant.
\end{center}
\del			decompositions \cite{bl96}:
\BE
	16=8^v_{-1}+8^{\5s}_1,~~~~~ \5{16}=8^v_{-1}+8^{\5s}_1,
		~~~~~ 10=1_{-2}+8^s_0+1_2. 
\EE
l.h.s. (with $h=1$):
\BE
	(\underbrace{c,{c\atop R},{0\atop\5s}}_{Q=1},v), ~~~~~
	(\underbrace{a,{a\atop R},{0\atop s}}_{Q=-1},v), ~~~~~
	(\underbrace{c,{c\atop R},{v\atop\5s}}_{Q=2},0).
\EE
Safer: consider R sector

r.h.s.:
\BE
	(c,c,0,v) ~ \to ~ (R,R,\5s,s) 
\EE
Twists:
\BEA
	J_s^{tot}&=&J_s^X\ex J_s^F\ex J_s^{(2)}\ex  J_s^{(8)}, ~~~~ N=4d\\
	J_B&=&1\ex (J_s^F)^d\ex J_s^{(2)}\ex 1, ~~~~ N=4, J_B\not\in\ca_R\ni
		J_B^2=J_v^F\ex J_v^{(2)}\\
	J_{A}&=&1\ex1\ex V_v^{(2)}\ex V_v^{(8)}, ~~~~ N=2, J_A\not\in\ca_L\\
	J_{x8}&=&V_v^X\ex 1\ex1\ex V_v^{(8)}, ~~~~ N=2
\EEA
$R_{AB}=\2=X_{AB}$, $X_{BA}=0$ $\then$ 
$Q_B^{(R)}\equiv\a_A/2$, $Q_A^{(L)}\equiv\a_B/2$, all other are charges
integer in all sectors of the theory.
\enddel
	\def\NS{\hbox{NS}} \def\Rp{{\hbox{R$_+$}}} \def\Rm{{\hbox{R$_-$}}}
For a field $\Ph_{a,Ja}$ that is twisted by
\BE
	J=J_\GSO^{2n}J_A^\a J_B^{2\b-\r} J_\X^\g, \qquad %n=0,\ldots,M-1,\quad
	\a,\b,\g,\r=0,1
\EE
this leads to the following charge projections for the monodromy charges
\BE	
	Q_\GSO\equiv-\8\2Q_{U(1)}\equiv0%%\8{K-1\04}\r
	, ~~~~ Q_A\equiv\8\2\r,~~~~ Q_B\equiv\8{K-1\04}\r,~~~~ Q_C\equiv0,
\EE
or, equivalently, 
$\5Q_\GSO\equiv\5Q_A\equiv\5Q_C\equiv0$ and $\5Q_B\equiv\8\2\a+\frac{K-1}4\r$
modulo $1$.

The massless matter representations (chiral superfields) as well as possible 
gauge group extensions (vector superfields) can now be enumerated
straightforwardly. Space-time quantum numbers come from representations of 
the right-moving chiral algebra while the gauge group representations
follow from left-moving CFT quantum numbers. The correspondences have been 
worked out for $E_5=SO(10)$, $E_4=SU(5)$ and $E_3=SU(3)\times SU(2)$
by Blumenhagen and Wisskirchen \cite{bl95} (cf. their tables in section 6).
%
% In the present extension of 
For the case $SO(8)\ex U(1)\subseteq E_5$ the massless matter 
representations are assembled by the orbits of $\JGSO$ as follows,
%\cite{bl95}
\BE
	16=8^\sc_{-1}+8^v_1, ~~~~~ \5{16}=8^v_{-1}+8^\sc_1, ~~~~~ 
	10=1_{-2}+8^s_0+1_2,
\EE
where the subscripts denote the $U(1)$ charges. %% $Q\equiv-2Q_\GSO$ mod 2.

Only gauge-singlet representations can depend on non-topological information,
i.e. uncharged fields with $r=0$ and $h_{int}=1$ in eq. 
(\ref{model-dependent}).  All charged matter fields and non-abelian
gauge group extensions can hence be determined in terms of the data encoded in
the extended Poincar\'e polynomial of $\cc'$. Our construction can thus
be used for all Landau-Ginzburg orbifolds 
based on $N=2$ SCFTs of the form $\cc'\otimes \cf$ with a 
Fermat factor $\cf\sim \Ph^K$ with $K\in2\IZ+1$.

\section{Geometry %, (0,2) $\mathbf\sigma$ models, 
			and vector bundle data}		\label{sec:5}
							% weight doubling

Witten's gauged linear sigma model \cite{wi93} made it 
possible to construct  a large class of $(0,2)$ string vacua \cite{di94}.
The starting point is a supersymmetric abelian gauge theory
that leads in the Calabi-Yau phase to a $\sigma$ model 
described by an exact sequence (monad)				\VS-6
\BE								\VS-3
	0~\rightarrow~ V ~\rightarrow~ \bigoplus^{r+1}_{i=1}{\cal O}(n_{i})
	\stackrel{F_i}{~\rightarrow~} {\cal O}(m)~\rightarrow~ 0
\EE
defining a  bundle $V$ of rank $r$ 
over a complete intersection Calabi-Yau %variety 
$X$. $F_{i}$ are 
homogeneous polynomials of degrees $m-n_{i}$ not vanishing simultaneously 
on $X$. For weighted
projective ambient spaces we can write this data as		\VS-6
\BE							\VS-3
	V_{n_1\ldots,n_r+1}[m]~\longrightarrow
		~{\Bbb P}_{w_{1},\ldots , w_{N+4}}[d_1,\ldots,d_N],
\EE 
where $r=4,5$ corresponds to unbroken gauge 
groups $SO(10)$ and $SU(5)$, respectively.
The Calabi-Yau condition $c_1(X)=0$ and the condition $c_{1}(V)=0$, 
which guarantees %that the topological obstruction to 
the existence of spinors, %  vanishes, 
read								\VS-3
\begin{equation}						\VS-3
	           \sum d_l-\sum w_j= m-\sum n_{i}=0 \label{VBc1}
\end{equation}
and the cancellation of gauge anomalies
$ch_{2}(V)=ch_{2}(TX)$ with $ch_{2}=\frac{1}{2}c_{1}^{2}-c_{2}$ implies 
the quadratic diophantine constraint				\VS-3
\begin{equation}						\VS-3
	\sum%_{j=1}^{N-3}
	d_{l}^{2} -\sum%_{i=1}^{N+1}
		w_{j}^{2}	=	m^{2}-\sum%_{i=1}^{r+1} 
		n_{i}^{2}.\label{VBc2}
\end{equation}
For a Calabi-Yau hypersurface $W=0$ % of degree $d$ 
the choice of $m=d=\sum w_j$ with $n_{i}=w_{i}$ solves these equations and
$F_{i}=\partial_{i}W$ corresponds to the $(2,2)$ case.
\del
	In the special case of $F_{i}=\partial_{i}W$, where $W$ is the
transversal polynomial that defines $X$ in ${\Bbb P}_{w_{1},\ldots,w_{5}}$, 
we have a $(2,2)$ model with gauge group $E_{6}$;
for a generic choice of $F_{i}$ the bundle $V$ will 
be a stable deformation $(0,2)$ of the extension of ${\cal T}_{X}$ 
by $\cal O$ \cite{di94}.

We begin with the vector bundle data $V_{n_1,\ldots ,n_5}[m]$ 
which are associated with the rational weights $q_i=n_i/m$ with 
$m=\sum n_i$ of a transversal CY hypersurface in weighted 
${\Bbb P}^4$ which is suggested by the ${\Bbb Z}_m$ quantum symmetry 
\cite{va891} of the corresponding $(0,2)$ LG model that should correspond
to the ${\Bbb Z}_m$ symmetry that comes with the generalized GSO 
projection in the Gepner model. 
>From the example in \cite{bl95} we guess that the base CY
variety should be a complete intersection of codimension 2 in 
weighted ${\Bbb P}^5$ 
(note that because the superpotential of 
%the supersymmetric $U(1)$ gauge 
the theory one begins with has the form $\int d^{2}z\,d\theta^{-}\hspace{2mm}
\Sigma_{j}W_{j}(\Phi)+P\Lambda^{i}F_{i}(\Phi)$, where the $F_{i}$ do not 
vanish simultaneously on $W_{j}=0$ \cite{di94}; it is not necessary that 
$W_{j}$ be transversal. 	\enddel

The suggested CFT/geometry correspondence \cite{bl95} assosiates the
vector bundle $V_{1,1,1,1,1}[5]$ over $\IP_{1,1,1,1,2,2}[4,4]$ to the 
(0,2) cousin of the Gepner model $3^5$. Since the twist $J_B$ 
that defines the (0,2) model only acts on one of the Fermat factors
we expect that this is part of a larger picture, where the Gepner model
data directly translate into vector bundle data $V_{n_1,\ldots,n_5}[m]$
with $k_i=m/n_i-2$. For the base manifold the doubling of the respective
weight seems to correspond to the doubling of the order of the twist group
by the Bonn twist $J_B$ (as compared to the standard construction).
We hence make the ansatz 					\VS-3
\BE								\VS-3
	V_{n_1,\ldots,n_5}[m]	\to
       {\Bbb P}_{n_1,\ldots ,n_4,2n_5,w_6}[d_1,d_2],
\EE
i.e. 	$w_i=n_i$ for $i<5$ and $w_5=2n_5$, and impose (\ref{VBc1}) 
and (\ref{VBc2}) or					\VS-3
\BE							\VS-3
	d_1+d_2=m+n_5+w_6,\qquad d_1^2+d_2^2=m^2+3n_5^2+w_6^2.
\EE
It is quite non-trivial and encouraging that this non-linear system has a
general solution $w_6=(m-n_5)/2=d_1/2$ and $d_2=(m+3n_5)/2$. We hence
conjecture a correspondence between the (0,2) models defined in the
previous section with the Distler-Kachru models defined by the data\cite{kr95}
								\VS-3
\BE								\VS-3
	V_{n_1,\ldots,n_5}[m]	\to			\label{conj}
       {\Bbb P}_{n_1,\ldots ,n_4,2n_5,\frac{m-n_5}{2}}[m-n_5,(m+3n_5)/2].
\EE
The increase of the codimension of the Calabi-Yau may be interpreted as
providing an additional field of degree $w_6=d_1/2$ that generates %%describing
the twisted sectors for the $\IZ_2$ orbifolding due to $J_B$.

In the Calabi--Yau phase a toric approach to the resolution 
of singularities  appears to be most natural \cite{di96}.
%(a toric approach to the resolution of singularities 
%actually appears to be most natural, so that transversality of the 
%polynomials that define the base space will not be necessary).
%
%LG has beed discussed in the recent paper \cite{bl97}, so we focus on CY.
For the (2,2) model the Newton polytope $\D$ of a generic transversal 
degree $m$ polynomial is reflexive and
its polar polytope $\D^*$ provides a desingularization of the
hypersurface in the weighted projective space $\IP_{n_1,\ldots,n_5}$ 
\cite{ba94}. For the complete intersection (\ref{conj}) the 
Batyrev-Borisov construction \cite{ba95} suggests to consider the Minkowski
sum $\D=\D_1+\D_2$ of the Newton polytopes $\D_l$ of degree $d_l$ 
polynomials w.r.t. the weights $w_j$. If $\D$ is reflexive then a natural 
resolution of singularities can again be based on a triangulation of 
the fan over $\D^*$. A useful 
collection of tools and formulas for further studies of this class of
models can be found in a paper by Blumenhagen \cite{bl97}.

\del
$\Delta_{4}$ the Newton polytope 
of degree $m$ monomials on ${\Bbb P}_{n_1,\ldots ,n_4,n_5}$. This is
the polytope which the hyperplane  
\BE
	x_1n_1+\ldots+x_4n_4=n_5(K-x_5)\label{4}
\EE  
where $K=m/n_5$, cuts out from the intersection of the 
positive half spaces in ${\Bbb Z}^{5}$. 
Therefore $M$ will be a Calabi-Yau 
hypersurface in a toric variety which is a blowup of ${\Bbb P}_{n_1,
\ldots ,n_4,n_5}$. If we do the same thing for $V_{n_1,\ldots,n_5}[m]\to 
{\Bbb P}_{n_1,\ldots ,n_4,2n_5,\frac{m-n_5}{2}}[m-n_5,(m+3n_5)/2]$ that is
associated to the $(0,2)$ model we get a polytope as the
intersection of the positive half spaces and the hyperplane 
\BE
	y_1n_1+\ldots+y_4n_4=n_5(\8{3K+1\02}-2y_5-\8{K-1\02}y_6)\label{5}
\EE
in ${\Bbb Z}^{6}$, so that $0\le y_6\le3$. The above mentioned 
theorem, however, does not apply in this case and
we need a specific argument. 
%Nevertheless, we sketch some general points below.
\new
	This is fine since we need a Minkowski sum anyway.
\endnew
As pointed out above, we also need some nef partition of
the reflexive polytope. It is therefore suggestive to 
begin with Newton polytopes
of monomials of degrees $d_1=m-n_5=(K-1)n_5$ and $d_2=(m+3n_5)/2=(K+3)n_5/2$
defined by  
\BEA
	\D^{'}_5: &&	{y_1n_1+\ldots+y_4n_4\0n_5}=K-1-2y_5-{K-1\02}y_6,\\
	\D^{''}_5: &&	{y_1n_1+\ldots+y_4n_4\0n_5}={K+3\02}-2y_5-{K-1\02}y_6,
\EEA
(or some `modifications' thereof) and check
$\Delta_{5}=\Delta^{'}_{5}+\Delta{''}_{5}$ for the reflexivity.
Using $x_1,\ldots,x_4$ as coordinates for $\D_4$ and $y_1,\ldots,y_4,y_6$
for $\D_5$ we get the correspondence $x_5=2y_5+{K-1\02}y_6-{K+1\02}$ and
\BE
	{y_1n_1+\ldots+y_4n_4\0n_5}=
	K-x_5=\begin{cases}{3K+1\02}-2y_5 & y_6=0\\ {K+1}-2y_5 & y_6=1\\
		{K+3\02}-2y_5 & y_6=2\\ 2(1-y_5) & y_6=3\end{cases}
\EE
In particular, $y_i=1$ is the unique interior point and there is always the
vertex $(0,0,0,0,1,3)$.
$(K-1)/2$ is even/odd for $K\in4\IZ\pm1$, so that `$K-x_5$' can be odd only
for $K\in4\IZ-1$.
There is a vertex $(0,0,0,0,{K+1\02},1)$ for $K\in4\IZ-1$, ~
and $(0,0,0,0,{3K+1\04},0)$ for $K\in4\IZ+1$ (the case of \cite{bl96}).\\

	first step of the process outlined in \cite{di96} is to resolve the
	basis MF (p.12 bottom)

\new
\endnew

%%%%%%%%%%%%%%%%%%%%%%%%%%%%%%%%%%%%%%%%%%%%%%%%%%%%%%%%%%%%%%%%%%%%%%%%%%%%

\hrule

								\addrefs

\begin{center}
\begin{tabular}{||c|cccc\TVR50||} \hline\hline
(0,2) models	& total & Fermat & $K\in4\IZ+1$ & both \\\hline
$LG_4$~ (2390)	&  1021 & 159	 & 351	& 55	\\
$LG_5$~ (5165)	&  2198 & 61	 & 583	& 20	\\\hline
$W\IP^4$ (7555)	&  3219 & 220	 & 934	& 75	\\
\hline\hline\end{tabular}
\end{center}
$F$-theory: $CY_4\to B_3\supset \S_2\leftarrow CY_3$
			\verb+echo "8 1 1 1 1 2 2 d=4 4" | nef.x -f -Lv -m+
\enddel

\section{Conclusion}	\label{sec:6}\VS-3

We discussed the construction of a large class of heterotic (0,2) Gepner-type
models in terms of simple current techniques and their generalization to
Landau-Ginzburg models based on the topological information encoded by the 
extended Poincar\'e polynomial. Already without %further 
orbifolding the 7555
transversal potentials lead to 3219 models, 220 of which are of Fermat type.

For a large subclass of the potentials the mirrors of the (2,2) models can
be constructed as orbifolds \cite{Berglund:1991pp,Kreuzer:1994uc}.
In this case our analysis provides the ingredients
for an orbifold mirror construction also for the (0,2) version, thus 
explaining the mirror symmetry that has been observed in orbifold spectra
\cite{bl961,bl971}. While an algorithm for the construction of the mirror
orbifold is known also in the presence of discrete torsions \cite{odt}, it
would be interesting to find an explicit formula for the mirror orbifold
in group theoretical terms.

In addition to the phenomenological interest of heterotic models
it would be interesting to test the proposed identifications 
by comparing spectra in geometrical phases \cite{bl97} 
and Yukawa couplings at the Landau-Ginzburg points \cite{Melnikov:2009nh},
and to study generalizations with smaller gauge groups.

\smallskip

{\it Acknowledgements.} I would like to thank Ron Donagi and
Emanuel Scheidegger for helpful discussions.
This work is supported in part by the {\it Austrian Research Funds} 
FWF under grant Nr. P18679.

\VS-5

%%%	\newpage

\begin{appendix}[\hbox{Gepner\,\,models, 
			torus\,\,orbifolds\,\&\,mirror\,\,symmetry}]
\VS-5
In accord with the three weighted projective spaces $\IP_{111}[3]$,
$\IW\IP_{112}[4]$ and $\IW\IP_{123}[6]$ that admit a transversal CY equation
of degree $d=3,4,6$, there are three Gepner models with levels $k=(1,1,1)$, 
$k=(2,2,0)$ and $k=(4,1,0)$, and super\-potentials $W=X^3+Y^3+Z^3$, 
$W=X^4+Y^4+Z^2$ and $W=X^6+Y^3+Z^2$, 
respectively, that describe 2d tori. While the K\"ahler modulus is fixed at 
the Landau-Ginzburg point at a value that is consistent with the
$\IZ_d$ quantum symmetry originating in the GSO projection, the complex
structure deformation corresponds to a deformation of $W$ by $\l XYZ$.
At the Gepner point $\l=0$ the complex structure moduli are $\t=e^{2\p i/d}$,
where $e^{2\p i/3}$ and $e^{2\p i/6}$ are related by $\t\to\t+1$.

We focus on $\IZ_2\ex\IZ_2$ orbifolds, whose abelian 
extensions were recently classified and compared to free fermion models
by Donagi and Wendland \cite{DoWe}.
Since we want to realize the $\IZ_2$'s as symmetries of Gepner models we
consider $\IW\IP_{112}[4]$ and $\IW\IP_{123}[6]$, for which a 
phase rotation of the first homogeneous coordinate corresponds to a phase
rotation by $2\p/d$ of the flat double-periodic torus coordinate $z\in T^2$
(this can be checked by counting fixed points and orders of stabilizers). 
The $\mathbb Z_2$ orbifold $z\to-z$ hence corresponds to the phase 
symmetry $\r=\mathbb Z_2:1\,0\,0$ in both cases.

With the notation of \cite{DoWe} as subscript and the Hodge numbers as 
supersprict, the four inequivalent orbifolds by a $\mathbb Z_2\ex\mathbb Z_2$
twist group $G_T$ are $X_{0-1}^{51,3}$, $X_{0-2}^{19,19}$, $X_{0-3}^{11,11}$,
and $X_{0-4}^{3,3}$. They differ by the number of shifts  $z\to z+\2$ that
are included and we can choose the following generators,\cite{DoWe}
{	\def\2{{\textstyle\frac12}}
\BEA
\HS-5	X_{0-1}^{51,3~}:&&\hbox{\fns$\displaystyle	
	\th^{(1)}(z_1,z_2,z_3)=(-z_1,z_2,-z_3)	\atop\displaystyle
	\th^{(2)}(z_1,z_2,z_3)=(z_1,-z_2,-z_3)	$} %\hbox{Vafa-Witten}
\\\HS-5	X_{0-2}^{19,19}:&&\hbox{\fns$\displaystyle
	\th^{(1)}(z_1,z_2,z_3)=(-z_1,z_2,-z_3)~~\, 	\atop\displaystyle
	\th^{(2)}(z_1,z_2,z_3)=(z_1,-z_2,\2\!-\!z_3)	$}%& %\hbox{Schoen}
\\\HS-5	\textstyle
	X_{0-3}^{11,11}:&&\hbox{\fns$\displaystyle
	\th^{(1)}(z_1,z_2,z_3)=(-z_1,z_2\!+\!\2,-z_3)	\atop\displaystyle
	\th^{(2)}(z_1,z_2,z_3)=(z_1,-z_2,\2\!-\!z_3)~~	$}%& %\hbox{Enriques}
\\\HS-5	\textstyle
	X_{0-4}^{3,3~~}:&&\hbox{\fns$\displaystyle
	\th^{(1)}(z_1,z_2,z_3)=(z_1\!+\!\2,-z_2,-z_3)	~~~\atop\displaystyle
	\th^{(2)}(z_1,z_2,z_3)=(-z_1,z_2\!+\!\2,\2\!-\!z_3) $}
						%& %\hbox{Donagi-Wendland}
\EEA}%
%\centerline{Schoen: $\begin{cases}\th_1(z_1,z_2,z_3)=(-z_1,z_2,-z_3)\\
%		\th_2(z_1,z_2,z_3)=(z_1,-z_2,-z_3+\2)\end{cases}$
%	Enriques: $\begin{cases}\th_1(z_1,z_2,z_3)=(-z_1,z_2+\2,-z_3)\\
%		\th_2(z_1,z_2,z_3)=(z_1,-z_2,-z_3+\2)\end{cases}$}
Only $\mathbb P_{112}[4]$ admits a second independent $\mathbb Z_2$ action, 
namely $\s=\mathbb Z_2:1\,0\,1$, which has no fixed points and hence 
corresponds 
to a shift $z\to z+\2$ of order 2. 
The product $\r\circ\s=\mathbb Z_2:0\,0\,1$ also has 
4 fixed points and corresponds to the rotation $z\to\2-z$ about $z=\frac14$, 
which is equivalent to $\r$. For the 
realization of $X_{0-n}$
% $X_\Ph^{(3,51)}$, $X_S^{(19,19)}$ and $X_E^{(11,11)}$ 
in terms of Gepner models we hence need at least $n-1$ factors 
of $\mathbb P_{112}[4]$. 
This can be confirmed by computing the Hodge numbers with the program
package PALP \cite{PALP}. In a UNIX shell environment the required input 
data can be assembled as follows, % shown in table \ref{tab:palp}.

%\begin{table}
\VS-2 \HS4
\hbox{\vbox{\begin{flushleft}\fns\tt%\scriptsize\tt
Weight1="6  1 2 3 1 2 3 1 2 3 "
\\TorusQ1="/Z6:~1 2 3 0 0 0 0 0 /Z6:~0 0 0 1 2 3 0 0 0"
\\Weight2="12 2 4 6 2 4 6 3 3 6 " 
\\TorusQ2="/Z6:~1 2 3 0 0 0 0 0 /Z6:~0 0 0 1 2 3 0 0 0"
\\Weight3="12 2 4 6 3 3 6 3 3 6 " 
\\TorusQ3="/Z6:~1 2 3 0 0 0 0 0 /Z4:~0 0 0 1 1 2 0 0 0"
\\Weight4="4  1 1 2 1 1 2 1 1 2 "
\\TorusQ4="/Z4:~1 1 2 0 0 0 0 0 /Z4:~0 0 0 1 1 2 0 0 0"
\\X01="\$Weight1 \$TorusQ1 /Z2:~1 0 0 0 0 0 1 0 0 /Z2:~0 0 0 1 0 0 1 0 0"
\\X02="\$Weight2 \$TorusQ2 /Z2:~1 0 0 0 0 0 1 0 0 /Z2:~0 0 0 1 0 0 0 0 1"
\\X03="\$Weight3 \$TorusQ3 /Z2:~1 0 0 1 0 1 1 0 0 /Z2:~0 0 0 1 0 0 0 0 1"
\\X04="\$Weight4 \$TorusQ4 /Z2:~1 0 1 1 0 0 1 0 0 /Z2:~1 0 0 1 0 1 0 0 1"
%\\\verb+X04="$Weight4 $TorusQ4 /Z2: 1 0 1 1 0 0 1 0 0 /Z2: 1 0 0 1 0 1 0 0 1"+
\\\verb+echo -e "$X01 \n$X02 \n$X03 \n$X04" | poly.x -lf+
\end{flushleft}}}

\def\MyCaption#1{
\VS-4 \fns\HS6	\stepcounter{MyTable} Table {\theMyTable}: \makeatletter
\immediate\write\@auxout{\string\newlabel{tab:palp}{\theMyTable}}
\makeatother\\[-8.2pt]\HS46
\parbox{92mm}{#1}\VS-55	}
%\end{table}
\VS-8

\noindent
%	PALP\cite{PALP} input for computing the Hodge data of the torus
%	orbifolds $X_{0-1}$ \ldots $X_{0-4}$. 
where	``{\tt Weight*}'' includes %provides
	a sufficient number of $\IP_{112}[4]$ factors for the shift
	symmetries, ``{\tt TorusQ*}'' provides two GSO projections for torus 
	factors (the overall GSO is automatic) and ``{\tt X0*}''
	completes the input line for the respective $\IZ_2\ex\IZ_2$ 
	orbifold $X_{0-1}$, \ldots, $X_{0-4}$. 
	The last line pipes the input into the executable {\tt poly.x}
	contained in PALP \cite{PALP},
	with flags ``-l'' and ``-f'' for ``Landau-Ginzburg'' and ``filter''
	(i.e. read input from pipe), respectively.

The mirror models can now be constructed using the Green-Plesser orbifold
construction. In \cite{DoWe} it was observed that discrete torsions
often provide the mirrors. This is special to $\IZ_2$-torsions, however,
for which a discrete torsion between two %generators 
phase symmetries of even order of the LG superpotential
can be switched on/off by redefinition of the action on massive fields $Z^2$,
as has been discussed in detail in \cite{odt}. For general orders of the
generators, the mirror models of orbifolds with discrete torsion again
have discrete torsion \cite{odt} and we do not know of
any indications that mirror symmetry and discrete torsion are related
for $\IZ_n$ twists with $n\neq2$ \cite{ade,lgt}.

In the classification of extensions $G_S\to G\to G_T$ of the 
twist group \cite{DoWe}, $G_S$ is the subgroup of shifts.
%, which contains additional shifts of order $2$. 
Only $\IP_{112}[4]$ admits a symmetry that corresponds to a
second independent shift $\s'$ of order 2, which however cannot
be diagonalized simultaneously with $\s$. It exchanges $X$ and $Y$ and 
reverses the sign of $Z$. The mirror construction in this case proceeds by
first taking the Green-Plesser mirror for the diagonal subgroup and then
performing the mirror moddings of the remaining twists on the mirror CFT, 
which may involve quantum symmetries. It would be interesting to use examples
from \cite{DoWe} with non-trivial fundamental groups to further
test the conjecture that mirror symmetry exchanges torsion in 
$H^2(X,\IZ)$ with torsion in $H^3(X,\IZ)$ \cite{BKtor}.

\end{appendix}	\bigskip	\def\2{{\frac12}}
\VS-5

\begin{appendix}[N=2 SCFT, simple currents \& minimal models]

				\label{app:N=2}	\VS-5
						\let\ket=\rangle
\def\nibf#1:{\noindent{\bf #1:}}

% \[	c=1 \hbox{ free boson at radius $R=2p/q$, } ~ h=\8{k^2\02N} 
%	\hbox{ with $N=2pq$ sectors: ~ } -\8{N\02}<k\le\8{N\02}		\]
%% P.Furlan, R.R. Paunov, I.T.Todorov, {\it Extended U(1) conformal
%% field theories and Z(K) parafermions,} Fortschr. Phys. 40 (1992) 211;\\
%% P. Furlan, G.M.Sotkov, Ya.Stanev, {\it Minimal models of U(1) conformal 
%% current algebra,} \jmp 29 (1988) 2311;\\
%% R.R.Paunov, I.T.Todorov, {\it Modular invariant QFT models of U(1) 
%%	conformal current algebra,} \plb 196 (1987) 519;\\
%% Buchholz, Mack, Todorov,  {\it in} Nucl. Phys. Proc. 5B (1988) 20

The $N=2$ superconformal algebra \cite{lvw} is generated by the Fourier 
modes of $T(z)$, of its fermionic superpartners $G^\pm(z)$, 
and of a $U(1)$ current $J(z)$
% W.Lerche, C.Vafa, N.Warner, NPB 324(89)427
\BEA  							\label{GrGs}
	&\{G_r^-,G_s^+\}=2L_{r+s}-(r-s)J_{r+s}+\8{c\03}(r^2-\8{1\04})\d_{r+s},&
\\	& [L_n,G_r^\pm]=(\8{n\02}-r)G_{n+r}^\pm,
	\qquad				~	[J_n,G_r^\pm]=\pm G_{n+r}^\pm,&
\\	& [L_n,J_m]=-mJ_{m+n},\qquad\quad	[J_m,J_n]=\8{c\03}m\d_{m+n},
\EEA
where $r,s\in\IZ+\2$ in the NS sector. According to (\ref{GrGs}) the Ramond
gound states $G_0|\a\rangle_R=0$ have $h_\a=c/24$. The analogous unitarity
bound in the NS sector is saturated by the chiral primary fields \cite{lvw} 
$G_{-\2}^+|\Ph\ket=0$, which obey $\{G_{\2}^-,G_{-\2}^+\}|\Ph\ket=
(2L_0-J_0)|\Ph\ket=0$ and hence $h=Q/2$. Their conjugate anti-chiral states
saturate the BPS bound $h=-Q/2$.

The N=2 algebra admits the continous spectral flow
\BE \textstyle\hbox{\small$
        L_n\rel {\cu_\th} \longrightarrow       %\cu_\th^{-1}L_n\cu_\th=
        L_n+\th J_n+{c\06}\th^2\d_n,~~~~~
    J_n\rel {\cu_\th} \longrightarrow   %\cu_\th^{-1}J_n\cu_\th=
        J_n+{c\03}\th\d_n,~~~~~
        G_r\rel {\cu_\th} \longrightarrow%    \cu_\th^{-1}G_r^\pm\cu_\th=
                G_{r\pm\th}^\pm$} 
\EE
which for $\th=\pm\2$ maps Ramond ground states into chiral and antichiral
primary fields, respectively. Spectral flow is best understood by 
bosonization of the $U(1)$ current $J(z)=i\sqrt{c/3}\,\6X(z)$ in terms of 
a free field $X$. A charged operator $\co_q$ can thus be written as a 
normal ordered product of a
vertex operator with a neutral operator $\co_0$, 		\vspace{-3pt}
\BE									\VS-3
	\co_q=e^{i\sqrt{3/c}\,qX}\,\co_0(\6X,\ldots,\ps,\ldots)
\EE
% $\cu_\th(z)\Ph_q(z',\5z')\sim(z-z')^{q\th}\Ph_{q+c\th/3}$;
The contribution of the vertex operator to $h$ is $3q^2\02c$ so that 
in unitary theories the
maximal charges of Ramond ground states and chiral primary states are 
$c/6$ and $c/3$, respectively. In particular, the Ramond ground state 
$J_\sp=e^{i\sqrt{c/12}\,X}$ with maximal charge $c/6$ is a 
% vertex operator, and hence a 
simple current. A short calculation shows that its monodromy charge 
is $Q_\sp=-\frac12Q$. If the $U(1)$ charges $Q$ are quantized in units
of $1/M$ in the NS sector then $c=3k/M$ for some integer $k$. Since
the $U(1)$ charges are shifted by $-c/6=-k/2M$ in the Ramond sector
the order $N_\sp$ of
$J_\sp$ is $2M$ if $k\in2\IZ$ and $4M$ if $k\not\in2\IZ$.

Already for $N=1$ SCFTs the supercurrent $G$ is a universal simple 
current, which we denote by $J_\su=G$. Its monodromy charge is $Q_\su=0$ 
for NS fields and $Q_\su=1/2$ for Ramond 
fields since $h_v=3/2$ and the conformal weights of superpartners differ by 
integers in the Ramond sector and by half-integers for NS states.
\del
$\ct=J+\th^+G^-+\th^-G^++\th^+\th^-T$; $D^\pm=Q^\pm=\6_{\th^\mp}+\th^\pm\6_z$,
$J=\th^+\6_{\th^+}-\th^-\6_{\th^-}$, $H=2\6_z$\\
R: $D^\pm=G_0^\pm$, $H=2L_0-c/12$; NS: $D^\pm=G_{-\2}^\pm$, $H=2L_{-1}$:\\
$\{Q^+,Q^-\}=H$, $[J,Q^\pm]=\pm Q^\pm$, $[H,J]=[H,Q^\pm]=0$

spectral flow (A.Schwimmer, N.Seiberg, PLB 184(87)191):
\BE \cu_\th^{-1}L_n\cu_\th=L_n+\th J_n+{c\06}\th^2\d_n,~~~
     \cu_\th^{-1}J_n\cu_\th=J_n+{c\03}\th\d_n,~~~
    \cu_\th^{-1}G_r^\pm\cu_\th=G_{r\pm\th}^\pm \EE
locality: $J(z)=\sqrt{c/3}\6\Ph(z)$, $\Ph_q=e^{i\sqrt{3/c}q\Ph}
\co_0(\6\Ph,\ps_0,\ldots)$\\
$\cu_\th(z)\Ph_q(z',\5z')\sim(z-z')^{q\th}\Ph_{q+c\th/3}$;
contribution of $q$ to $h$ is $3q^2\02c$.\\
space-time susy: A.Sen, NPB 278(86)289; NPB 284(87)423; \\
T.Banks, L.J.Dixon, D.Friedan, E.Martinec, NPB 299(88)613; \\
T.Banks, L.Dixon, NPB 307(88)93

unitarity constraints (W.Boucher, D.Friedan, A.Kent, PLB 172 (1986) 191):\\
left chiral: $G_{-\2}^+|\Ph\>=0$, primary $\then\{G_{\2}^-,G_{-\2}^+\}|\Ph\>=
(2L_0-J_0)|\Ph\>=0$, $h=q/2$\\
as $G_{-\2}^+=(G_\2^-)^\hc$ unitarity $\then h\ge|q|/2$; 	% vice versa:
in turn: \\
$h=q/2\then\<\Ph|\{G_{\2}^-,G_{-\2}^+\}|\Ph\>=
|G_\2^-|\Ph\>|^2+|G_{-\2}^+|\Ph\>|^2=0$. $J_n|\Ph\>=0$ for $n>0$\\
(as $h\ge|q|/2$ and $[L_0,J_n]=-mJ_n$),
thus $G_{n+\2}^\pm=\pm[J_n,G_\2^\pm]$ vanishes for $n\ge0$, i.e. primary.\\
OPE: $\Ph(z)\c(z')\sim(z-z')^{h_{\ph\c}-h_\Ph-h_\c}:\Ph\c:$,
$h_{\ph\c}\ge\2(q_\Ph+q_\c)=h_\Ph+h_\c$\\
``Hodge decomposition'': minimize $\left||\Ph_0\>=|\Ph\>-G_\2^-|\Ph_+\>-
G_{-\2}^+|\Ph_-\>\right|^2=0$ $\then$ $|\Ph_0\>$ chiral primary, as
$\<\Ph_0|(|\Ph_0\>-G_\2^-|\e_+\>-G_{-\2}^+|\e_-\>)=0$ ($|\Ph\>$ chiral $\then
|\Ph_+\>=0$)\\
$\{G_{3/2}^-,G_{-3/2}^+\}=2L_0-3J_0+2c/3$ and $h=q/2$ impies $h\le c/6$.\\
R: $h\ge c/24$ (as for $N=1$), $h=c/24$ contribute to index
$\tr(-)^F$ (Witten, NPB202(82)253)\\
for $q_L-q_R\in\IZ$: $\tr(-)^F=e^{i\p(J_0-\5J_0)}$, anticommutes with $G^\pm$
and squares to $1$. LR-sym. flow:\\
$\tr(-)^F\equiv\tr_R(-)^{(J_0-\5J_0)}q^{L_0-c/24}\5q^{\5L_0-c/24}=
\sum_{(c,c)}e^{i\p(q_l-q_r)}=\tr_{\ch_\th}\left((-)^{J_0^\th-\5J_0^\th}
q^{L_0^\th-c/24}\5q^{\5L_0^\th-c/24}\right)$
\BE P(t,\5t)\equiv\tr_{(c,c)}t^{J_0}\5t^{\5J_0}
	     =(t\5t)^{c\06}\tr_Rt^{J_0}\5t^{\5J_0}{}_{|_{g.s.}}\EE
charge conjugation in R $\then$ $P(t,\5t)=(t\5t)^{c\03}P(1/t,1/\5t)$
(Poincar\'e duality).\\
$J=i\sqrt{c\03}\6\Ph$, $\co_{q_L,q_R}=e^{i\sqrt{3\0c}(q_L\Ph_L-q_R\Ph_R)}
\c_{0,0}$, $\cu_\th=e^{-i\th\sqrt{c\03}(\Ph_L-\Ph_R)}$,
$({c\03},{c\03})$-state: $\r=\cu_{-1}|0\>$.\\
integral left charge $\then c=3D$, $(-)^{F_L}=e^{i\p J_0}$,
$\cu_{(\2,0)}=e^{-{i\02}\sqrt{c\03}\Ph_L}$: (R,R)$\to$(NS,R)\\
(\& similar flow in ghost and space-time sector -- imposing GSO to states
with definite $(-)^{F_L}$).\\
$\r_L=e^{i\sqrt{c/3}\Ph_L}$, $\r=\r_L\otimes\r_R$; $e^{-i\sqrt{c/3}\Ph_L}$:
(c,c)$\to$(a,c)\\
$P_{(a,c)}=\tr_{(a,c)}t^{J_0}\5t^{\5J_0}$, $P_{(c,c)}=\sum b_{p,q}t^p\5t^q
\then P_{(a,c)}=\sum b_{d-p,q}t^p\5t^q=P_{(c,a)}$ (Hodge duality)\\
conjecture: $W=G_{-{3\02}}^+\5G_{-{3\02}}^+$

\nibf Bosonization of $U(1)$ current: \\		\def\sp{s}\def\su{v}
simple current with conformal weight $3/2$ and order 2, the
supercurrent $J_\su$, follows already from $N=1$ SUSY; its 
monodromy charge is $0$ for NS states and $1/2$ in the Ramond sector. 
For $N=2$ SUCO models we have at least one more simple current, namely  
the Ramond ground state
$J_\sp$ of highest $U(1)$ charge, which in our normalization is $c/6$: 
{\it spinor current}, implements the spectral flow: In terms of the
bosonized $U(1)$ current of the $N=2$ algebra $J(z)=\sqrt{c\03}\6X(z)$ it is 
given by
$J_\sp = \exp{i\sqrt{3\0c}X}$. It follows from the operator product of the
free boson $X$ that the monodromy charge $Q_\sp$ of the spinor current
is related to the $U(1)$ charge $Q$ by $Q_\sp\equiv-Q/2$ modulo 1.
Thus, if $1/M$ is the charge quantum in the NS sector and $\hat c=c/3=k/M$, 
then the order of the spinor current $J_{\sp}$ is $2M$ if $k$ is even and 
$4M$ if $k$ is odd, because the charges in the Ramond sector are shifted by 
$c/6$. If $k$ is odd, then $J_{\su}=(J_{\sp})^{2M}$.

$J_{\sp}$ is a Ramond ground state, hence $h_{\sp}=c/24$ and
$Q_{\sp}(J_{\sp})=-c/12$. 
\enddel
Putting the pieces together we find the matrix of monodromies
									\VS-7
\BE	\label{MMMM}							\VS-7
     	R_{\su,\su}=0, ~~~~ R_{\su,\sp}=1/2, ~~~~
%		R_{\sp,\sp}=1-\d_\IZ(k/4)-c/12,
	R_{\sp,\sp}=n-c/12~~\hbox{with $n=$\fns\scriptsize$
	\begin{cases}0&k\in4\IZ\medskip\\[-7pt]1&k\not\in4\IZ\end{cases}$}
%	\begin{cases}k/2&k~\rm even\medskip\\[-7pt]~1&\,k~\rm odd\end{cases}$}
\EE	\nobreak
where we used $h_{\sp}=c/24$ and $Q_{\sp}(J_{\sp})=-c/12$. 
%In the case of \nn\ minimal models $k$ is the level and $M=k+2$.
Note that $J_\sp^{2M}=J_\su^k$		(since the monodromy charges agree) 
so that the universal
center is $\IZ_{2M}\times\IZ_2$ for $k\in2\IZ$ and $\IZ_{4M}$
for $k\not\in2\IZ$.

\goodbreak
			\VS-5
\section{$N=2$ minimal models}		\def\TS{\TVR50}	\def\q{m}	
			\VS-5
				\label{app:MM}

Minimal models have a number of different realizations. Here we use the
coset construction for the $N=2$ superconformal series $\cc_k$	\VS-3
\BE								\VS-5
	(SU(2)_k\ex U(1)_4)/U(1)_{2K},\quad~~	c=3k/K ~\hbox{ with }~ K=k+2
\EE
as a quotient of $SU(2)$ level $k$ for $k\in\IN$ times $U(1)_4\cong
SO(2)_1$ by $U(1)_{2K}$. Primary fields $\Ph_\q^{ls}$ are labelled accordingly
by $0\le l\le k$, $s\mod4$ and $m\mod 2K$ with the branching rule
$l+\q+s\in2\IZ$. The fusion rules are				\VS-3
\BE								\VS-5
	\Ph_{\q_1}^{l_1s_1}\ex\Ph_{\q_2}^{l_2s_2}
	=\sum_{l=|l_1-l_2|}^{\min(l_1+l_2,k)-|k-l_1-l_2|} 
	\Ph_{\q_1+\q_2}^{l,s_1+s_2}
\EE
so that $\Ph_\q^{0s}$ and $\Ph_{\q+K}^{k,s+2}$ are simple currents.
The conformal weights and the $U(1)$ charges obey			\VS-3
\BE	\textstyle							\VS-3
	h\equiv{l(l+2)-\q^2\04K}+{s^2\08}~\mod~1, ~~~
	Q\equiv{s\02}-{\q\0K}~\mod~2		~~~~	
	\hbox{exact for}	~\bigl\{\!{|\q-s|\le l\atop \,-1\le s\le 1}\;
\EEL {hN2MM}
where the NS and R sectors correspond to even and odd $s$, respectively.
The formulas (\ref{hN2MM}) are exact in the standard range $|\q-s|\le l$, 
$-1\le s\le 1$ 
%%			which involves a choice of chirality (see below), 
and otherwise sufficient to determine the monodromy charges
of simple currents. In particular, the selection rule $l+m+s\in2\IZ$ is
implemented by integrality of the monodromy charge $Q_K^{k2}$ of the
simple current $\Ph_K^{k2}$, which has integral conformal weight.
According to the rules for modular invariance the branching rule thus
necessitates the field identification				\VS-3
\BE								\VS-3
	\Ph_\q^{ls}\sim\Ph_{\q+K}^{k-l,s+2} \quad\hbox{with}\quad 
		J_{id}=\Ph_K^{k2},\quad	Q_{id}\equiv (l+\q+s)/2
\EE
due to an extension of the chiral algebra by the 
``identification current''~$J_{id}$.
%%%						chiral for $D_{2n},E_6,E_8$
The center of the minimal model at level $k$ is hence of order $4K$ and
generated by the spectral flow current 
$J_\sp:=\Ph_1^{01}\sim\Ph_{1-K}^{k3}$ 
% (inverse spectral flow operator) $J_s:=\Ph_1^{01}$ and the
and the supercurrent $J_\su:=\Ph_0^{02}\sim\Ph_K^{k0}$
with $J_\sp^{2K}=J_\su^k$; more generally
all above formulas for $N=2$ SCFTs apply with $M=K$.
\del

Since $J_s^{2K}=J_v^K$ and $J_v^2=\Id$ it is isomorphic to $\IZ_{2K}\ex\IZ_2$
for $k$ even and to $\IZ_{4K}$ for $k$ odd. Using (\ref{hN2MM}) we find the
following monodromy charges:
\\[6pt]
$J_{s}=\Ph_1^{01}$~~~~$Q_{s}=-Q/2$~~~$N_{s}=\{{2K{\rm~ even}\atop 4K
	{\rm~~odd}}$ ~~~$Q_{s}(J_{s})\equiv R_{ss}=n-{k\04K}$ ~~with  
	$n\equiv k{2K\0N_{s}}\equiv
		\{{0{\rm~for~}k\in4\IZ\atop1{\rm~for~}k\not\in4\IZ}$
	%	$\{{k/2\hbox{ even}\atop ~k\hbox{ ~~odd }}$
\\[4pt]
$J_{v}=\Ph_0^{02}$~~~~$Q_{v}=~s/2$~~~~~$N_{v}=~2$~~~~~~~~~~~~$Q_{v}(J_{v})
	\equiv R_{vv}=1$ ~~~~~~~~~$Q_{v}(J_{s})=R_{vs}=1/2$ 
% The monodromies $Q_s$ of the supercurrent show that R and NS sector 
% correspond to $s$ odd and even, respectively. 
\enddel
Ramond ground states and (anti)chiral primary fields are now easily 
identified as follows,
\\[3pt]				%identified as follows,
\centerline{\small
\begin{tabular}{||c|c|c||} \hline\hline	\VR{3.7}{1.6}	
	anti-chiral primary		%~$\VR42  G^-_{-\frac12}|\Ph\ket=0 $ 
	& %\atop h=-Q/2\VR32
	Ramond ground states 	%~$\VR42 G_0^\pm|\Ph\ket=0 $ 
	& %\atop h=c/24\VR32
	chiral primary %~$\VR42 G^+_{-1/2}|\Ph\ket=0$
\\ \hline%\atop h=Q/2\VR32
\VR42	$\Ph_{l}^{l0}\sim\Ph_{K+l}^{k-l,2} ~\to~ |l\ket_a$	& 
   $\Ph_{\pm(l+1)}^{l,\pm1}\sim\Ph_{\mp(k-l+1)}^{k-l,\mp1}~\to~|l_\pm\ket_R$ &
	$\Ph_{-l}^{l0}\sim\Ph_{K-l}^{k-l,2}~\to~ |l\ket_c$\\%\hline
\VR42	$Q=-{l\0K}$, ~~ $h=-{Q\02}$ & 
	$Q=\pm({c\06}-{l\0K})$, ~~~ $h={c\024}$ 	&
	$Q={l\0K}$, ~~~ $h={Q\02}$	\\
\hline\hline\end{tabular}}
\\[3pt]
The Landau-Ginzburg description of the minimal model with the diagonal
modular invariant has superpotential $W=X^K$ with $X\sim\Ph_{-1}^{1,0}$.

%
% The primary fields $\Ph_\q^{ls}$ correspond to characters of the (bosonic
% part of) the chiral algebra, i.e. they contain $U(1)$ descendents and 
% bilinears of supersymmetry creation operators. 
In order to determine the conformal weights and multiplicities of all 
fields relevant for massless string spectra we follow the discussion in ref.
\cite{sc90} and first note that the supercurrent $J_v$ acts as
$J_\su\Ph_\q^{ls}=\Ph_\q^{l,s+2}\sim\Ph_{\q\pm K}^{k-l,s}$. Choosing $\q$ 
such that $-K<\q\le K$ we find that $\q\to \q-K$ for $\q>0$ and 
$\q\to \q+K$ for $\q\le0$. It is then straightforward to check 
that %%	$|\q|\to K-|\q|$ and 
$l+1-|\q|\to -(l+1-|\q|)$, i.e. the fields inside 
the cone $|\q|\le l+1$ are mapped to the outside and vice versa.

In the NS sector we choose $s=0$. Then (\ref{hN2MM}) gives the correct
value of $h$ inside the cone, i.e. for $|\q|\le l$. The conformal weight
of the respective superpartner is $h+\2$ and its multiplicity is $2$ unless
$G_{-1/2}^+$ or $G_{-1/2}^-$ vanishes. This happens for $|\q|=l$, for 
which the multiplicity of the superpartner is 1 for $l>0$. For $l=\q=0$, i.e.
the superpartner $J_v$ of the identity,
the lowest states have $h=3/2$ with multiplicity 2. 
% (the two supercurrents contained in $J_v$).

{\footnotesize
%$|\Ph\ket={\{G_0^+,G_0^-\}\02h-c/12}|\Ph\ket=G_0^+(~)+G_0^-(~)$ ~~~
%	$[J_0,G^\pm]=\pm G^\pm$, $G^{(1)}=G^++G^-$, $iG^{(2)}=G^+-G^-$:
%	$G^{(i)}\in[J_v]$.
}

In the R sector highest weight states are annihilated by $G_0^+$ or $G_0^-$.
They thus come in pairs $\Ph_\q^{l,\pm1}$ that are related by the action of 
$G_0^\pm$. Usually we can fulfill $|\q|<l$ by field identification, in which
case $h$ is degenerate and given correctly by (\ref{hN2MM}). 
The only exception is $|\q|=l+1$ where $G_0^+=G_0^-=0$. In that case one has
to make a choice of chirality: The Ramond ground states have $h=c/24$ in
accordance with (\ref{hN2MM}), and their superpartners have $h=1+c/24$. The
choice $\q=l+1$ and $s=1$ leads to the standard range given in (\ref{hN2MM}). 
The only descendent that plays a role for the massless spectrum of strings
is the descendent $J_{-1}|0\ket$ of the vacuum.
% coming from a spin 1 chiral 
% field. The $N=2$ minimal models have 1 such field, the $U(1)$ current.

%  $R_{gs}$ [$h={c\024}$]: $\Ph_{\pm(l+1)}^{l,\pm1}\sim\Ph_{\mp(k-l+1)}^{k-l,
% 	\mp1}$, $NS_{c}$ [$Q=2h$]: $\Ph_{-l}^{l0}\sim\Ph_{K-l}^{k-l,2}$,
%  	$NS_{a}$ [$Q=-2h$]: $\Ph_{l}^{l0}\sim\Ph_{K+l}^{k-l,2}$\\
%$Q_{s}\Ph_{-l-1}^{l,-1}=({c\012}-{l\02K})|l\>_R$ ~~ 
%  $Q_{s}\Ph_{-l}^{l0}=-{l\02K}|l\>_c$ ~~ $Q_{s}\Ph_{l}^{l0}={l\02K}|l\>_a$ 

\end{appendix}

\Count			\bigskip
\begin{appendix}[Matter representations and gauge groups]

r.h.s.:
\BE
	|\underbrace{R,R,\5s}_{Q=-1},\5s\ket 
\EE

l.h.s. (with $h=1$):
\BE
	|\underbrace{R,{R\atop a},{\5s\atop0}}_{Q=-1},\5s\ket, ~~~~~
	|\underbrace{R,{R\atop c},{s\atop 0}}_{Q=1},\5s\ket, ~~~~~
	|\underbrace{R,{R\atop c,a},{s,\5s\atop0}}_{Q=0},s\ket.
\EE
The restriction to $\5s$ ($s$) for $16$ ($\5{16}$) comes from disregarding the
universal state with extremal R ground state charge. In the NS sector only
the ground state can contribute since $h$ from spinor of $SO(8)$ together with
R ground state in $X$ already is larger than 1/2.

\del

r.h.s.:
\BE
	|\underbrace{R,R,\5s}_{Q=-1},\5s\ket 
\EE

l.h.s. (with $h=1$):
\BE
	|\underbrace{R,{R\atop a},{\5s\atop0}}_{Q=-1},\5s\ket, ~~~~~
	|\underbrace{R,{R\atop c},{s\atop 0}}_{Q=1},\5s\ket, ~~~~~
	|\underbrace{R,{R\atop c,a},{s,\5s\atop0}}_{Q=0},s\ket.
\EE
The restriction to $\5s$ ($s$) for $16$ ($\5{16}$) comes from disregarding the
universal state with extremal R ground state charge. In the NS sector only
the ground state can contribute since $h$ from spinor of $SO(8)$ together with
R ground state in $X$ already is larger than 1/2.
We need to get from the l.h.s. to the r.h.s. with the following twists:
\BEA
	J_\GSO	% =J_s^{tot}
		&=&J_s^\X\ex J_s^F\ex J_s^{(2)}\ex J_s^{(8)}, 
		~~~~ N=4M\\
	J_B&=&1\ex (J_s^F)^K\ex J_s^{(2)}\ex 1, ~~~~ N=4, J_B\not\in\ca_R\ni
		J_B^2=J_v^F\ex J_v^{(2)}\\
	J_{A}&=&1\ex1\ex J_v^{(2)}\ex J_v^{(8)}, ~~~~ N=2, J_A\not\in\ca_L\\
	J_{\X}&=&J_v^\X\ex 1\ex1\ex J_v^{(8)}, ~~~~ N=2
\EEA
\enddel

For

{	\scriptsize

$\2=X_{AB}$, $X_{BA}=0$ $\then$ 
	$\5Q_B\equiv\a_A/2$, $Q_A\equiv\a_B/2$, all other charges must be
integer in all sectors of the theory.
\\[5pt]
{\bf Projection:} $Q_A\equiv \a_B/2$ o.k., 
$Q_B\equiv0\equiv KQ_s^{F}+Q_s^{(2)}$ (i.e. the projection can be implemented
by having all alignments -- except for the case of $\a_B$ twist -- and 
projecting to integral $Q_{GSO}$ and $Q_B$ on the left).
\BEA
	R: &&Q_B(|x,l,\5s,\5s\rangle)=K(-\8\2)(l-k/2)/K+\8{1\04}
		=\8{K-1\04}-\8{l\02}\in\IZ\\
	NS: &&Q_B(|x,l,0,\5s\rangle)=\pm l/2=\8{K+1\04}\in\IZ
\EEA
For the various representations we thus need twists
\BEA
	16~\Rm:	&&	J_\GSO^{2n} ~ J_\X^\x ~ J_A^\a ~ J_B^{2\b}~~~	
	~~~~ \a+\b\eq0~~~ \a+\x\eq0 ~~~\then\s=+	\\
	16~\NS:	&&	J_\GSO^{2n} ~ J_\X^\x ~ J_A^\a ~ J_B^{2\b-1}	
	~~~~ \a+\b\eq0~~~ \a+\x\eq0 ~~~\then\s=+	~~n_F=n_\X+\8{K+1\02}
	~~~~~~
\EEA
\BEA
	\5{16}~\Rp:&&	J_\GSO^{2n} ~ J_\X^\x ~ J_A^\a ~ J_B^{2\b}~~~
	~~~~ \a+\b\eq1~~~ \a+\x\eq0 ~~~\then\s=-	\\
	\5{16}~\NS:&&	J_\GSO^{2n} ~ J_\X^\x ~ J_A^\a ~ J_B^{2\b-1}	
	~~~~ \a+\b\eq0~~~ \a+\x\eq0 ~~~\then\s=+	~~n_F=n_\X+\8{K-1\02}
	~~~~~~
\EEA
\BEA
	10~\Rp:&&	J_\GSO^{2n} ~ J_\X^\x ~ J_A^\a ~ J_B^{2\b}~~~	
	~~~~ \a+\b\eq1~~~ \a+\x\eq1 ~\then\s=+		\\
	10~\Rm:&&	J_\GSO^{2n} ~ J_\X^\x ~ J_A^\a ~ J_B^{2\b}~~~	
	~~~~ \a+\b\eq0~~~ \a+\x\eq1 ~\then\s=-		\\
	10~\NS_+:   &&	J_\GSO^{2n} ~ J_\X^\x ~ J_A^\a ~ J_B^{2\b-1}	
	~~~~ \a+\b\eq0~~~ \a+\x\eq1 ~\then\s=-	~~n_F=n_\X+\8{K-1\02}\\
	10~\NS_-:   &&	J_\GSO^{2n} ~ J_\X^\x ~ J_A^\a ~ J_B^{2\b-1}	
	~~~~ \a+\b\eq0~~~ \a+\x\eq1 ~\then\s=-	~~n_F=n_\X+\8{K+1\02}
	~~~~~~~~~~~
\EEA
Here $0\le n<M$ if $M\in2\IZ$ and $0\le n<2M$ if $M\not\in2\IZ$; in the
latter case $J_\GSO^{2M}=J_v^\X\ex J_v^F\ex J_v^{(2)}$ and we can sum over
$0\le n<M$ if we add the modulus of contributions of both signs in the EPP
... $(R,R)\to(J_vR,J_vR)$ ...

For the alignment currents we have $\a+\x\equiv 0 \mod 2$ for spinors and 
$\a+\x\equiv 1 \mod 2$ for the vector; the presense of all the alignment
currents as possible twists is the reason for the definition of EPP: To
write down the modular invariant in terms of $X=\2R+E$ we need to choose a
basis, but in practice we only need to count the total number of 
supercurrents that we need for a shift modulo two and then now that an 
appropriate choice for, in our case, $\a,\b,\x$ does or does not exist.

What we will need is the number of Ramond ground states (or superpartners
of these) with certain left/right
charges at a certain location of the orbit of $J_s^{2n}$ (with an additional
$J_s^{\pm1}$ if we consider (anti)chiral states instead of Ramond ground
states). This is exactly the information that is encoded in the EPP: ...

}	% \end tiny

\noindent	{\bf	Relation to (2,2):} 
no $Q_B$ twist and projection. Then $\#10=\#16+\#\516$ from spectral flow?

%	\newpage

Compute charges and sign and projections
\\	since $(-)^s=(-)^n$ in the untwisted Fermat case only even $n$ in
	all cases; if $M$ is odd then we keep both signs.

	\scriptsize

\begin{center}

      \def\LR#1#2{$\hbox{\small$#1$}\VR{4.5}2\atop\VR3{2.5}\hbox{\small$#2$}$}
      \def\Lr#1#2{\LR{#1}{\hbox{\scriptsize$#2$}}}

{\fns	\begin{tabular}{||c|c||c|c|r||c\TVR52||} \hline\hline
\multicolumn2{||c||}{\large\bf16}	& $n=n_0+sK$ & $Q_X'=Q_X+{c_X\06}$ 
	& $\s_X\cdot\s_F=\s$ ~ & $Q_B\in\IZ$\\ \hline\hline
$|R,R,\sc,\sc\ket$ & $n\in K\IZ$ & $0\le l\le K-2$ 
	& \LR {Q'_X=1-l/K} {\5Q'_X=1-l/K} & $1\cdot(-)^s=+1$   
	& $l\in{K-1\02}+2\IZ$		     			\\ \hline
    & $n\not\in K\IZ$ & \LR {l=n_0-1~~~~~~} {\5l=K-n_0-1}
	& \LR {Q'_X=1-{n_0-1\0K}} {\5Q'_X={n_0+1\0K}~~~~~} 
	& $-1\cdot(-)^{n+1}=+1$ & $n_0\in{K+1\02}+2\IZ$		\\ \hline\hline
$|R,a,0,\sc\ket$ 
    & \Lr {n'\in K\IZ} { n'=n-{K+1\02}} & 
	\Lr {l=\5l=n_0} {n_0={K+1\02}} 
	& \LR {Q'_X={K+1\02K}} {\5Q'_X={K+3\02K}} & $-1\cdot(-)^{n'}=+1$   
	& ${K-1\02}\in2\IZ$		     			\\ \hline
    & \Lr {n\in K\IZ}  { n_0'\neq0\then n_0=0} & 
	\Lr {l=n_0'-1} { n_0'=n_0-{K+1\02}}
	& \LR {Q'_X={K+1\02K}} {\5Q'_X={K+1\02K}} & $1\cdot(-)^n=+1$   
	& ${K-1\02}\in2\IZ$		     			\\ \hline\hline
\end{tabular}}
\\[5mm]		
{\bf Table I:} $X$-sector charges for Fermat-sector states $|l,\5l\ket$, 
	sign $\s$ in EPP, and 	\\projection $Q_B=Q_s^{(2)}-{K\02}Q_F\in\IZ$
	for generations.	\HS29

\bigskip

{\scriptsize	\begin{tabular}{||c|c||c|c|r||c\TVR52||} \hline\hline
\multicolumn2{||c||}{\large$\bf{\5{16}}$} & $n=n_0+sK$ & $Q_X'=Q_X+{c_X\06}$ 
	& $\s_X\cdot\s_F=\s$ ~ & $Q_B\in\IZ$\\ \hline\hline
$|R,R,s,\sc\ket$ & $n\in K\IZ$ & $0\le l\le K-2$ 
	& \LR {Q'_X=2-l/K} {\5Q'_X=1-l/K} & $-1\cdot(-)^n=-1$   
	& $l\in{K+1\02}+2\IZ$		     			\\ \hline
    & $n\not\in K\IZ$ & \LR {l=n_0-1~~~~~~} {\5l=K-n_0-1}
	& \LR {Q'_X=2-{n_0-1\0K}} {\5Q'_X={n_0+1\0K}~~~~~} 
	& $1\cdot(-)^{n+1}=-1$ & $n_0\in{K-1\02}+2\IZ$		\\ \hline\hline
$|R,c,0,\sc\ket$ 
    & \Lr {n'\in K\IZ} { n'=n-{K-1\02}} & 
	\Lr {l=\5l=n_0} {n_0={K-1\02}} 
	& \LR {Q'_X={K+1\02K}} {\5Q'_X={K+3\02K}} & $-1\cdot(-)^{n'}=+1$   
	& ${K-1\02}\in2\IZ$		     			\\ \hline
    & \Lr {n\in K\IZ}  { n_0'\neq0\then n_0=0} & 
	\Lr {l=n_0'-1} { n_0'=n_0-{K-1\02}}
	& \LR {Q'_X={K+1\02K}} {\5Q'_X={K+1\02K}} & $1\cdot(-)^n=+1$   
	& ${K-1\02}\in2\IZ$		     			\\ \hline\hline
\end{tabular}}
\\[5mm]
{\bf Table II:} $X$-sector charges for Fermat-sector states $|l,\5l\ket$, 
	sign $\s$ in EPP, and 	\\projection $Q_B=Q_s^{(2)}-{K\02}Q_F\in\IZ$
	for anti-generations.	\HS5

\bigskip

{\scriptsize	\begin{tabular}{||c|c||c|c|r||c\TVR52||} \hline\hline
\multicolumn2{||c||}{\large\bf10}	& $n=n_0+sK$ & $Q_X'=Q_X+{c_X\06}$ 
	& $\s_X\cdot\s_F=\s$ ~ & $Q_B\in\IZ$\\ \hline\hline
$|\underbrace{R,R,s}_{Q=0},s\ket$ & $n\in K\IZ$ & $0\le l\le K-2$ 
	& \LR {Q'_X=1-l/K} {\5Q'_X=1-l/K} & $1\cdot(-)^n=+1$   
	& $l\in{K+1\02}+2\IZ$		     			\\ \hline
    & $n\not\in K\IZ$ & \LR {l=n_0-1~~~~~~} {\5l=K-n_0-1}
	& \LR {Q'_X=1-{n_0-1\0K}} {\5Q'_X={n_0+1\0K}~~~~~} 
					& $-1\cdot(-)^{n+1}=+1$ 
	& $n_0\in{K-1\02}+2\IZ$					\\ \hline\hline
$|\underbrace{R,R,\sc}_{Q=0},s\ket$ & $n\in K\IZ$ & $0\le l\le K-2$ 
	& \LR {Q'_X=2-l/K} {\5Q'_X=1-l/K} & $-1\cdot(-)^n=-1$   
	& $l\in{K-1\02}+2\IZ$		     			\\ \hline
    & $n\not\in K\IZ$ & \LR {l=n_0-1~~~~~~} {\5l=K-n_0-1}
	& \LR {Q'_X=2-{n_0-1\0K}} {\5Q'_X={n_0+1\0K}~~~~~} 
	& $1\cdot(-)^{n+1}=-1$ & $n_0\in{K+1\02}+2\IZ$		\\ \hline\hline
$|\underbrace{R,c,0}_{Q=0},s\ket$ 
    & \Lr {n'\in K\IZ} { n'=n+{K+1\02}} & 
	\Lr {l=\5l=n_0} {n_0={K-1\02}} 
	& \LR {Q'_X={K+1\02K}} {\5Q'_X={K+3\02K}} & $-1\cdot(-)^{n'}=+1$   
	& ${K-1\02}\in2\IZ$		     			\\ \hline
    & \Lr {n\in K\IZ}  { n_0'\neq0\then n_0=0} & 
	\Lr {l=n_0'-1} { n_0'=n_0+{K+1\02}}
	& \LR {Q'_X={K+1\02K}} {\5Q'_X={K+1\02K}} & $1\cdot(-)^n=+1$   
	& ${K-1\02}\in2\IZ$		     			\\ \hline\hline
$|\underbrace{R,a,0}_{Q=0},s\ket$ 
    & \Lr {n'\in K\IZ} { n'=n+{K+1\02}} & 
	\Lr {l=\5l=n_0} {n_0={K-1\02}} 
	& \LR {Q'_X={K+1\02K}} {\5Q'_X={K+3\02K}} & $-1\cdot(-)^{n'}=+1$   
	& ${K-1\02}\in2\IZ$		     			\\ \hline
    & \Lr {n\in K\IZ}  { n_0'\neq0\then n_0=0} & 
	\Lr {l=n_0'-1} { n_0'=n_0+{K+1\02}}
	& \LR {Q'_X={K+1\02K}} {\5Q'_X={K+1\02K}} & $1\cdot(-)^n=+1$   
	& ${K-1\02}\in2\IZ$		     			\\ \hline\hline
\end{tabular}}
\\[5mm]
{\bf Table III:} $X$-sector charges and projection for Fermat-sector states 
		$|l,\5l\ket$ for SO(10) vectors.	\HS29

\end{center}

\BE
	\a+\x=\begin{cases}0 & spinor\\ 1 & vector 10\\\end{cases} ~~~~
	\a+\b+n=\begin{cases}0 & |R,*,0/\sc,*\ket\\ 1 & |R,R,s,*\ket
			\end{cases} ~~~~
	\r=\begin{cases}0 & R / aligned\\ 1 & NS/no-align\end{cases}
\EE

\BE
	\s_\x^{(n)}\cdot\s_F^{(n')}=\s=(-)^{\b+\x}, ~~~~~
	n'=n-\r\8{K\pm1\02} \hbox{ for }
	\begin{cases}a&(16,10)\\ c&(\516,10)\end{cases}
\EE

i.e. at twist $(J_s^\X)^{2n}(J_s^F)^{2n'}$.

\del

\newpage

\begin{tabular}{||c|c||c|c|r||c\TVR52||} \hline\hline
\multicolumn2{||c||}{\bf10}	& $n=n_0+sK$ & $Q_X'=Q_X+{K+1\0K}$ 
	& $\s_F\cdot\s_X=\s$ ~ & $Q_B\in\IZ$\\ \hline\hline
$|R,R,s,s\ket$ & $n\in K\IZ$ & $0\le l\le K-2$ 
	& \LR {Q'_X=1-l/K} {\5Q'_X=1-l/K} & $(-)^s\cdot 1=+1$   
	& $l\in{K+1\02}+2\IZ$		     			\\ \hline
& $n\not\in K\IZ$ & \LR {l=n_0-1~~~~~~} {\5l=K-n_0-1}
	& \LR {Q'_X=1-l/K} {\5Q'_X=1-l/K} & $(-)^s\cdot 1=+1$   
	& $l\in{K+1\02}+2\IZ$		     			\\ \hline\hline
$|R,R,\sc,s\ket$ & $n\in K\IZ$ & $0\le l\le K-2$ 
	& \LR {Q'_X=1-l/K} {\5Q'_X=1-l/K} & $(-)^s\cdot 1=-1$   
	& $l\in{K+1\02}+2\IZ$		     			\\ \hline
& $n\not\in K\IZ$ & \LR {l=n_0-1~~~~~~} {\5l=K-n_0-1}
	& \LR {Q'_X=1-l/K} {\5Q'_X=1-l/K}  & $(-)^s\cdot 1=-1$   
	& $l\in{K+1\02}+2\IZ$		     			\\ \hline
$|R,c,0,s\ket$ & $n\in K\IZ$ & $0\le l\le K-2$ 
	& \LR {Q'_X=1-l/K} {\5Q'_X=1-l/K} & $(-)^s\cdot 1=-1$   
	& $l\in{K+1\02}+2\IZ$		     			\\ \hline
& \Lr {n'\in K\IZ} { n'=n+{K-1\02}} & 
	\LR {l=n_0-1~~~~~~} {\5l=K-n_0-1}
	& \LR {Q'_X=1-l/K} {\5Q'_X=1-l/K} & $(-)^s\cdot 1=-1$   
	& $l\in{K+1\02}+2\IZ$		     			\\ \hline\hline
\hline
\end{tabular}
\\[5mm]
	Table III: 	Charges for F-states, Supercurrent-twist, 
			$Q_B=Q_s^{(2)}-{K\02}Q_F$.

\enddel

\del
${16}$ 	&  ~~~~ $\8{Q_B\equiv0}\atop n=n_0+s\,K$ ~~~~ & keep sign $+$ $=(-)^n$
	~\,	& 	$(J_s^{tot})^{2n}$ ~~$\bigl\{{0\le n<M~~~M\in2\IZ\atop
			0\le n < 2M~~M\not\in2\IZ}$ 
\\\hline\hline
R: $n\in K\IZ$	&	$l\in 2\IZ+{K-1\02} \atop\VR33	0\le l\le K-2$ 
		&	$(-)^n=(-)^s\s_X$
		&	${	Q'_X=1-l/K\atop	\VR43	\5Q'_X=1-l/K }$
\\\hline
R: $n\not\in K\IZ$ &  $n_0\in2\IZ+{K+1\02}\atop l=n_0-1, \5l=K-n_0-1$ 
		& 	$(-)^n=(-)^{n+1}\s_X$
		&  	${  Q'_X=1-{n_0-1\0K}\atop \VR43 \5Q'_X={n_0+1\0K}}$
\\\hline
NS: $n\in K\IZ$	&	$n'_0\neq0~~\then \atop n'_0-1=  l={K-1\02}\in2\IZ$
		& 	$(-)^n=(-)^n\s_X$
	  	&  	${Q'_X={K+1\02K}\atop \VR43 \5Q'_X={K+1\02K}}$
\\\hline
NS: $n'\in K\IZ$ & $n_0=l={K-1\02}\in2\IZ\atop	n'=n+{K+1\02}$
		& 	$(-)^n=(-)^{n'}\s_X$
		&  	${Q'_X={K+1\02K}\atop \VR43 \5Q'_X={K+3\02K}}$
\\\hline\hline\end{tabular}
\enddel

\newpage	%  Weyl = 16 ... usual convention for (anti) generations

%	Use $Q'=Q+c/6$ to simplify and conform with EPP; 
%	$\s_x=(-)^{Q_x-\5Q_x+n\hat c_x} ~~\then~~ \5Q_x-Q_x\equiv n{2\0K}$.

\noindent				
	$Q({s\atop\sc})=\pm\8\2$  ~ ~ 
	$Q_X'=Q_X+{1\06}c_X=Q_X+{K+1\0K}$ (like epp-convention). ~ ~
	(2,2) model: $\bigl\{$${\rm keep~all}~Q_B~~\atop\rm no~`NS'~sector$
\\[2pt]
	$\s_x=(-)^{Q_x-\5Q_x+n\hat c_x} ~\then~ \5Q_x-Q_x\equiv n{2\0K}$ ~ ~ 
	$J_B=(J_s^F)^KJ_s^{(2)} ~\then~ Q_B=-{K\02}Q_{XF}+Q_s^{(2)}.$
\\[4pt]
	Twist $n=n_0+sK$: ~$0\le n<${\footnotesize$
	\begin{cases}M&\HS-3 even\\2M&odd\end{cases}$}	
		with good signs only; ~~
	$Q'=l/K$ ~  $\bigl\{{l=n_0-1~~~\atop \5l=K-1-n_0}$.
\\[9pt]
	{\bf Weyl spinors}: right ${16}\to 8_{-1}^\sc = |R,R,\sc,\sc\rangle$,
	$~\5Q=\5Q_{XF}-\8\2=-1 ~~\then~~ \5Q'_X=1-\5Q'_F$
\\[5pt]
	{\bf Generations} from {\bf NS} sector: 
	$|R,R,\sc,\sc\rangle_{\rm right}=
	J|R,a,0,\sc\rangle_{\rm left}~\then$ twist 
	$J=(J_s^F)(J_s^{(2)})^{-1}=(J_s^F)^{K+1}J_B^{-1}$
		$~\then~n':=n_F=n_X+{K+1\02}$.  
	~~$Q_F=-{l\0K}={1-K\02K}$ $~\then~Q_X'={K+1\0K}-1-Q_F={K+1\02K}$.
	\\untwisted: $\5Q_F'=Q_F+\hat c_F={K-3\02K}$, ~~
	twisted: $\5Q_F'={K-2\0K}-{K-3\02K}={K-1\02K}$ ~...~ $\5Q_X'=1-\5Q_F'$
\\[9pt]
	{\bf R anti-generations}: left $\5{16}\to8_{1}^\sc=|R,R,s,\sc\rangle$,
	$~Q=Q_{XF}+\8\2=1 ~~\then~~ Q'_X=2-Q'_F$,
\\[5pt]
	{\bf NS anti-generations}: 
	$|R,R,\sc,\sc\rangle_{\rm right}=J|R,c,0,\sc\rangle_{\rm left}~\then$ 
	twist $J=(J_s^F)^{-1}(J_s^{(2)})^{-1}=(J_s^F)^{K-1}J_B^{-1}$
		$~\then~n':=n_F=n_X+{K-1\02}$.
	~~$Q_F'=Q_F={l\0K}={K-1\02K}$ $~\then~Q_X'={K+1\0K}+1-Q_F={3K+3\02K}$.

\begin{center}
\begin{tabular}{||l|l|ll\TVR52|l||} \hline\hline
${16}_R$ & $|R,R,\sc,\sc\rangle$	&~ $Q_B\equiv{1\04}-{K\02}{l-k/2\0K}$
	& $\then ~~l\in{K-1\02}+2\IZ$		& $Q'_X=1-Q'_F$\\
${16}_{NS}$ & $|R,a,0,\sc\rangle$	&~ $Q_B\equiv l/2$	
	& $\then ~~l={K-1\02}\in 2\IZ$	& $Q'_X={K+1\02K}$\\
$\5{16}_R$ & $|R,R,s,\sc\rangle$	&~ $Q_B\equiv{3\04}-{K\02}{l-k/2\0K}$
	& $\then ~~l\in{K+1\02}+2\IZ$ 	& $Q'_X=2-Q'_F$ \\
$\5{16}_{NS}$ & $|R,c,0,\sc\rangle$	&~ $Q_B\equiv-l/2$	
	& $\then ~~l={K-1\02}\in 2\IZ$	& $Q'_X={3K+3\02K}$	\\
\hline\hline
\end{tabular}
\end{center}

{\VS-17	\tiny
\[ 1 + 4tT + 10{t^2}{T^2} + 20{t^3}{T^3} + 31{t^4}{T^4} + 
   40{t^5}{T^5} + 44{t^6}{T^6} + 40{t^7}{T^7} + 
   31{t^8}{T^8} + 20{t^9}{T^9} + 10{t^{10}}{T^{10}} + 
   4{t^{11}}{T^{11}} + {t^{12}}{T^{12}} + {T^{12}}\,x + 
   {t^4}\,{T^8}\,{x^2} + {t^8}\,{T^4}\,{x^3} + {t^{12}}\,{x^4}	\VS-27	
\]
}

\begin{center}
\begin{tabular}{||c|c|l|c\TVR74||} \hline\hline
${16}$ 	&  ~~~~ $\8{Q_B\equiv0}\atop n=n_0+s\,K$ ~~~~ & keep sign $+$ $=(-)^n$
	~\,	& 	$(J_s^{tot})^{2n}$ ~~$\bigl\{{0\le n<M~~~M\in2\IZ\atop
			0\le n < 2M~~M\not\in2\IZ}$ 
\\\hline\hline
R: $n\in K\IZ$	&	$l\in 2\IZ+{K-1\02} \atop\VR33	0\le l\le K-2$ 
		&	$(-)^n=(-)^s\s_X$
		&	${	Q'_X=1-l/K\atop	\VR43	\5Q'_X=1-l/K }$
\\\hline
R: $n\not\in K\IZ$ &  $n_0\in2\IZ+{K+1\02}\atop l=n_0-1, \5l=K-n_0-1$ 
		& 	$(-)^n=(-)^{n+1}\s_X$
		&  	${  Q'_X=1-{n_0-1\0K}\atop \VR43 \5Q'_X={n_0+1\0K}}$
\\\hline
NS: $n\in K\IZ$	&	$n'_0\neq0~~\then \atop n'_0-1=  l={K-1\02}\in2\IZ$
		& 	$(-)^n=(-)^n\s_X$
	  	&  	${Q'_X={K+1\02K}\atop \VR43 \5Q'_X={K+1\02K}}$
\\\hline
NS: $n'\in K\IZ$ & $n_0=l={K-1\02}\in2\IZ\atop	n'=n+{K+1\02}$
		& 	$(-)^n=(-)^{n'}\s_X$
		&  	${Q'_X={K+1\02K}\atop \VR43 \5Q'_X={K+3\02K}}$
\\\hline\hline\end{tabular}
\end{center}

\begin{center}
\begin{tabular}{||c|c|l|c\TVR74||} \hline\hline
$\5{16}$ 	& ~~~~ $\8{Q_B\equiv0}\atop n=n_0+s\,K$ ~~~~ 
		& 	keep sign $-$ $=(-)^{n+1}$
		& 	$(J_s^{tot})^{2n}$ ~~$\bigl\{{0\le n<M~~~M\in2\IZ\atop
			0\le n < 2M~~M\not\in2\IZ}$ 
\\\hline\hline
R: $n\in K\IZ$	&	$l\in 2\IZ+{K+1\02} \atop\VR33	0\le l\le K-2$ 
		&	$(-)^{n+1}=(-)^n\s_X$
		&	${	Q'_X=2-l/K\atop	\VR43	\5Q'_X=1-l/K }$
\\\hline
R: $n\not\in K\IZ$ &  $n_0\in2\IZ+{K-1\02}\atop l=n_0-1, \5l=K-n_0-1$ 
		& 	$(-)^{n+1}=(-)^{n+1}\s_X$
		&  	${  Q'_X=2-{n_0-1\0K}\atop \VR43 \5Q'_X={n_0+1\0K}}$
\\\hline
NS: $n\in K\IZ$	&	$n'_0\neq0~~\then \atop n'_0-1=  l={K-1\02}\in2\IZ$
		& 	$(-)^n=(-)^n\s_X$
	  	&  	${Q'_X={K+1\02K}\atop \VR43 \5Q'_X={K+1\02K}}$
\\\hline
NS: $n'\in K\IZ$ & $n_0=l={K-1\02}\in2\IZ\atop	n'=n+{K+1\02}$
		& 	$(-)^n=(-)^{n'}\s_X$
		&  	${Q'_X={K+1\02K}\atop \VR43 \5Q'_X={K+3\02K}}$
\\\hline\hline\end{tabular}
\end{center}

\hrule\bigskip

If we want to count the number of fermion generations and anti-generations
we have to be careful in order not to overcount: Of course we get, for each
positive helicity $SO(10)$-spinor a negative chirality SO(10)-antispinor by
charge conjugation. In addition, all building blocks of the left-moving 
representations of the gauge group are combined with all right-moving
building blocks of the $SO(10)$-representations that is mapped to the
representations of the space-time Lorentz group by the bosonic string map,
as is quaranteed by the appropriate extensions of the chiral algebra by
simple currents.

16 Weyl spinor: 
\BE
	16\to 	\HS37 		\HS37 \otimes \HS37	\5{16}\to
\EE

\del
$\5{16}$ Weyl spinor: 
\BE
	\5{16}\to 	\HS37 		\HS37 \otimes \HS37	\5{16}\to
\EE

10 Weyl spinor:
\BE
	10\to 	\HS37 		\HS37 \otimes \HS37	\5{16}\to
\EE
\enddel

Instead of looking for a Weyl spinor we could look for a space-time scalar, 
which is related to a vector of $SO(10)$ by the bosonic/fermionic string map, 
i.e. for example ...

$SO(2)$ charge of $s, \sc, v, 0$ mod 2 
(see Gepner \cite{Gepner} Triest Lectures: Table I / p.17).
\BE
	Q_s^{tot}\equiv Q_s^{(8)}+Q_s^{(2)}-\2 Q^{c=9}_{U(1)} ~~~~\to~~~
	Q=Q^{c=9}_{U(1)}-2 Q_s^{(2)}=Q_{Gepner}
\EE
Puzzle: why are $8^\sc_{\pm1}$ different?

Working hypothesis: if $Q=1$ only spinor of $SO(2)$ contributes.

Right-moving: alignment $\then$ need $R \ex R\ex s\ex \sc$ with 
$Q(R\ex R)=1/2$ for Weyl fermion.

Left-moving 16:  in NS of $SO(2)$ ...
$v$ not allowed by $h=1$; in $R$ sector $\sc$ not 
allowed by working hypothesis.\\
In $R$ sector as usual: need $Q=1$ for $R\ex R\ex s\ex \sc$. In $NS$:
\BE	\HS-35
	\hat c_X=3-\hat c_F(K)=2\8{K+1\0K} ~~\then~~ h_l=\8{l\02K} ~~\then~~ 
	1=h_{tot}=h_X+h_l+0+\8\2, ~h_X=\hat c_X/8 ~~\then~~ l=\8{K-1\02}.
\EE
same for $\5{16}$ with $s\to\sc$, $Q\to-Q$ and $chiral\to anti-chiral$ in 
Fermat factor.\\
For $10$ we have same with $Q=0$; $v$ still excluded by $h=1$, but now
$R\ex R\ex s/\sc\ex \sc$ and $R\ex c/a\ex0\ex \sc$.

Sector $n$ / twist by $(J_s^{tot})^{2n}$:  $~\#(16)=N_R+N_{NS}$ ~(Bonn twist
even/odd in $N_R/N_{NS}$; $J_B^2$ is alignment as everything else except 
$J_s^{tot}$.
\BE
	N_R=\sum_n c_n\left({Q=\2-{l-k/2\0K}\atop\5Q=\2-{\5l-k/2\0K}}
			\right)_{|_{l(n),\5l(n)}},
\EE
\BE
	N_{NS}=\sum_n c_n\left(\8{Q(l(K))\atop\5Q(\5l(K,n))}
			\right)_{|_{l(n),\5l(n)}},
\EE

In twisted sectors of MM $exp(t)$ plus $exp(\5t)$ is constant, equal to $k=K-2$
\BE
	\cp^{(MM)}(x;t^K,\5t^K)=\sum_{l=1}^{K-1}(t\5t)^{l-1}
		{1-(-x)^l~\5t^{K-2l}	\0	1-(-x)^K}
	={P(t\5t) - \sum_{l=1}^{K-1}(-x)^lt^{l-1}\5t^{K-1-l} \0 1-(-x)^K}
\EE
with the ordinary Poincar\'e polynomial $P(t)={1-t^{K-1}\01-t}$.

Keeping the $\5t$-dependence makes the sign redundant:
\BE
	\5Q\equiv Q+l\8{c\03}-k ~\mod2 ~~~~ \hbox{ in sector } J_s^lJ_v^k
	~~~~ \then ~~~ k=Q-\5Q+l\8{c\03}
\EE

{\footnotesize
decompositions:
\BE
	16=8^\sc_{-1}+8^v_1,~~~~~ \5{16}=8^v_{-1}+8^\sc_1,~~~~~ 
	10=1_{-2}+8^s_0+1_2. 
\EE

l.h.s. (with $h=1$):
\BE
	(\underbrace{c,{c\atop R},{0\atop\5s}}_{Q=1},v), ~~~~~
	(\underbrace{a,{a\atop R},{0\atop s}}_{Q=-1},v), ~~~~~
	(\underbrace{c,{c\atop R},{v\atop\5s}}_{Q=2},0).
\EE
Safer: consider R sector\\
r.h.s.:
\BE
	(c,c,0,v) ~ \to ~ (R,R,\5s,\5s) 
\EE
}

Twists:
\BEA
	J_s^{tot}&=&J_s^X\ex J_s^F\ex J_s^{(2)}\ex  J_s^{(8)}, ~~~~ N=4d\\
	J_B&=&1\ex (J_s^F)^d\ex J_s^{(2)}\ex 1, ~~~~ N=4, J_B\not\in\ca_R\ni
		J_B^2=J_v^F\ex J_v^{(2)}\\
	J_{A}&=&1\ex1\ex V_v^{(2)}\ex V_v^{(8)}, ~~~~ N=2, J_A\not\in\ca_L\\
	J_{x8}&=&V_v^X\ex 1\ex1\ex V_v^{(8)}, ~~~~ N=2
\EEA
$R_{AB}=\2=X_{AB}$, $X_{BA}=0$ $\then$ 
	$Q_B^{(R)}\equiv\a_A/2$, $Q_A^{(L)}\equiv\a_B/2$, all other are charges
integer in all sectors of the theory.

{\footnotesize
{\bf Projection:} $Q_A\equiv \a_B/2$ o.k., 
$Q_B\equiv0\equiv KQ_s^{F}+Q_s^{(2)}$
\BEA
	R: &&Q_B(|x,l,\5s,\5s\rangle)=K(-\8\2)(l-k/2)/K+\8{1\04}
		=\8{K-1\04}-\8{l\02}\in\IZ\\
	NS: &&Q_B(|x,l,0,\5s\rangle)=\pm l/2=\8{K-1\04}\in\IZ
\EEA
}

\del
Weyl fermions / right $\5{16}=8_1^\sc$: ~$|R,R,s,\sc\rangle$ 
	$\5Q=\5Q_{XF}+\8\2=1 \then \5Q'_X=2-\5Q'_F$ (since $Q(s)=\8\2$).

\bigskip\hrule\noindent
anti-gen. / left $\5{16}$: R like above $Q'_X=2-Q'_F$
~~ ($|R,R,\sc,\sc\rangle$ requires max. charge/gauginos)\\
NS: $|R,c,0,\sc\rangle$
~~$Q_X'=Q_X+\hat c_X/2={3\02}{K+1\0K}$ since $l={K-1\02}$ and $Q=l/K$;
anti-chiral forbidden by charge, $v^{(2)}$ forbidden by conformal weight.

\bigskip\hrule\noindent
generations / left ${16}=8_{-1}^\sc$: $|R,R,\sc,\sc\rangle$ 
	$Q=Q_{XF}-\8\2=-1 ~\then~ Q'_X=1-Q'_F$ (since $Q(\5s)=-\8\2$).\\
NS: $|R,a,0,\sc\rangle$
~~$Q_X'=-1+{K+1\0K}-Q_F={K+1\02K}$ since $Q_X+Q_F=-1$ and $Q_F=-{K-1\02K}$.

\bigskip\hrule\noindent
\begin{center}
\begin{tabular}{||l|l|ll\TVR52|l||} \hline\hline
$\5{16}_R$ & $|R,R,s,\sc\rangle$	&~ $Q_B\equiv{3\04}-{K\02}{l-k/2\0K}$
	& $\then ~~l\in{K+1\02}+2\IZ$ 	& $Q'_X=2-Q'_F,~\5Q'_X=2-\5Q'_F$ \\
$\5{16}_{NS}$ & $|R,c,0,\sc\rangle$	&~ $Q_B\equiv-l/2$	
	& $\then ~~l={K-1\02}\in 2\IZ$	& $Q'_X={3K+3\02K}$	\\
${16}_R$ & $|R,R,\sc,\sc\rangle$	&~ $Q_B\equiv{1\04}-{K\02}{l-k/2\0K}$
	& $\then ~~l\in{K-1\02}+2\IZ$		& $Q'_X=1-Q'_F$\\
${16}_{NS}$ & $|R,a,0,\sc\rangle$	&~ $Q_B\equiv l/2$	
	& $\then ~~l={K-1\02}\in 2\IZ$	& $Q'_X={K+1\02K}$
\\\hline\hline
\end{tabular}
\end{center}

{\VS-22	\tiny
\[ 1 + 4tT + 10{t^2}{T^2} + 20{t^3}{T^3} + 31{t^4}{T^4} + 
   40{t^5}{T^5} + 44{t^6}{T^6} + 40{t^7}{T^7} + 
   31{t^8}{T^8} + 20{t^9}{T^9} + 10{t^{10}}{T^{10}} + 
   4{t^{11}}{T^{11}} + {t^{12}}{T^{12}} + {T^{12}}\,x + 
   {t^4}\,{T^8}\,{x^2} + {t^8}\,{T^4}\,{x^3} + {t^{12}}\,{x^4}	\VS-22	
\]
}

\noindent
Use $Q'=Q+c/6$ to simplify and conform with EPP; 
	$\s_x=(-)^{Q_x-\5Q_x+n\hat c_x} ~~\then~~ \5Q_x-Q_x\equiv n{2\0K}$.
\begin{center}
\begin{tabular}{||c|c|l|c\TVR74||} \hline\hline
$\5{16}$ 	& ~~~~~~ $Q_B\equiv0$ ~~~~~~	& ~~~ sign ~~~ 
		&	$(J_s^{tot})^{2n}$ :: $0\le n < M$ 
\\\hline\hline
R: $n\in K\IZ$	&	$l\in 2\IZ+{K-1\02} \atop\VR33	0\le l\le K-2$ 
		&	$1=+\s_X$
		&	${	Q'_X=2-l/K\atop	\VR43	\5Q'_X=2-l/K }$
\\\hline
R: $n\not\in K\IZ$ &  $n_0={K-1\02}\in2\IZ\atop n_0=n\mod K$ 
		& 	$(-)^n=(-)^{n+1}\s_X$
		&  	${  Q'_X=2-{n_0-1\0K}\atop \VR43 \5Q'_X=1+{n_0+1\0K}}$
\\\hline
NS: $n\in K\IZ$	&	$l={K-1\02}\in2\IZ	$	
		& 	$???=+\s_X$
	  	&  	${Q'_X={3K+3\02K}\atop \VR43 \5Q'_X={3K+1\02K}}$
\\\hline
NS: $n\not\in K\IZ$ & $n_0-1={K-1\02}\in2\IZ\atop n_0=n\mod K$
		& 	$???=(-)^{n+1}\s_X$
		&  	${Q'_X={3K+3\02K}\atop \VR43 \5Q'_X={3K-1\02K}}$
\\\hline\hline\end{tabular}
\end{center}

\begin{center}
\begin{tabular}{||c|c|l|c\TVR74||} \hline\hline
${16}$ 	& ~~~~~~ $Q_B\equiv0$ ~~~~~~	& ~~~ sign ~~~ 
		&	$(J_s^{tot})^{2n}$ :: $0\le n < M$ 
\\\hline\hline
R: $n\in K\IZ$	&	$l\in 2\IZ+{K-1\02} \atop\VR33	0\le l\le K-2$ 
		&	$???=\s_X$
		&	${	Q'_X=1-l/K\atop	\VR43	\5Q'_X=2-l/K }$
\\\hline
R: $n\not\in K\IZ$ &  $n_0={K-1\02}\in2\IZ\atop n_0=n\mod K$ 
		& 	$???=(-)^{n+1}\s_X$
		&  	${  Q'_X=2-{n_0-1\0K}\atop \VR43 \5Q'_X=1+{n_0+1\0K}}$
\\\hline
NS: $n\in K\IZ$	&	$l={K-1\02}\in2\IZ	$	
		& 	$???=+\s_X$
	  	&  	${Q'_X={3K+3\02K}\atop \VR43 \5Q'_X={3K+1\02K}}$
\\\hline
NS: $n\not\in K\IZ$ & $n_0-1={K-1\02}\in2\IZ\atop n_0=n\mod K$
		& 	$???=(-)^{n+1}\s_X$
		&  	${Q'_X={3K+3\02K}\atop \VR43 \5Q'_X={3K-1\02K}}$
\\\hline\hline\end{tabular}
\end{center}

\del
\begin{center}
\begin{tabular}{||c|c|c|c\TVR74||} \hline\hline
16 $R::\s=0\atop NS::\s=1$	& ~~~~~~ $Q_B\equiv0$ ~~~~~~	& ~~~ sign ~~~ 
		&	twist $(J_s^{tot})^{2n}\ex (J_B)^\s\ex J_v$'s
								\\\hline\hline
R: $n\in K\IZ$	&	$l\in 2\IZ+{K-1\02} \atop\VR33	0\le l\le K-2$ & +
		&${	Q'=2-l/K
	\atop	\VR43	\5Q'=2-l/K				}$\\\hline
R: $n\not\in K\IZ$ &	&	&  			\\\hline
NS: $n\in K\IZ$	&	&  	&  			\\\hline
NS: $n\not\in K\IZ$ &	&	&  			\\\hline
\hline\end{tabular}
\end{center}
\enddel

\newpage	%  Weyl = 16 ... usual convention for (anti) generations

%	Use $Q'=Q+c/6$ to simplify and conform with EPP; 
%	$\s_x=(-)^{Q_x-\5Q_x+n\hat c_x} ~~\then~~ \5Q_x-Q_x\equiv n{2\0K}$.

\noindent				
	$Q({s\atop\sc})=\pm\8\2$  ~ ~ 
	$Q_X'=Q_X+{1\06}c_X=Q_X+{K+1\0K}$ (like epp-convention). ~ ~
	(2,2) model: $\bigl\{$${\rm keep~all}~Q_B~~\atop\rm no~`NS'~sector$
\\[2pt]
	$\s_x=(-)^{Q_x-\5Q_x+n\hat c_x} ~\then~ \5Q_x-Q_x\equiv n{2\0K}$ ~ ~ 
	$J_B=(J_s^F)^KJ_s^{(2)} ~\then~ Q_B=-{K\02}Q_{XF}+Q_s^{(2)}.$
\\[4pt]
	Twist $n=n_0+sK$: ~$0\le n<${\footnotesize$
	\begin{cases}M&\HS-3 even\\2M&odd\end{cases}$}	
	with good signs only; ~~
	$Q'=l/K$ ~  $\bigl\{{l=n_0-1~~~\atop \5l=K-1-n_0}$.
\\[9pt]
	{\bf Weyl spinors}: right ${16}\to 8_{-1}^\sc = |R,R,\sc,\sc\rangle$,
	$~\5Q=\5Q_{XF}-\8\2=-1 ~~\then~~ \5Q'_X=1-\5Q'_F$
\\[5pt]
	{\bf Generations} from {\bf NS} sector: 
	$|R,R,\sc,\sc\rangle_{\rm right}=
	J|R,a,0,\sc\rangle_{\rm left}~\then$ twist 
	$J=(J_s^F)(J_s^{(2)})^{-1}=(J_s^F)^{K+1}J_B^{-1}$
		$~\then~n':=n_F=n_X+{K+1\02}$.  
	~~$Q_F=-{l\0K}={1-K\02K}$ $~\then~Q_X'={K+1\0K}-1-Q_F={K+1\02K}$.
	\\untwisted: $\5Q_F'=Q_F+\hat c_F={K-3\02K}$, ~~
	twisted: $\5Q_F'={K-2\0K}-{K-3\02K}={K-1\02K}$ ~...~ $\5Q_X'=1-\5Q_F'$
\\[9pt]
	{\bf R anti-generations}: left $\5{16}\to8_{1}^\sc=|R,R,s,\sc\rangle$,
	$~Q=Q_{XF}+\8\2=1 ~~\then~~ Q'_X=2-Q'_F$,
\\[5pt]
	{\bf NS anti-generations}: 
	$|R,R,\sc,\sc\rangle_{\rm right}=J|R,c,0,\sc\rangle_{\rm left}~\then$ 
	twist $J=(J_s^F)^{-1}(J_s^{(2)})^{-1}=(J_s^F)^{K-1}J_B^{-1}$
		$~\then~n':=n_F=n_X+{K-1\02}$.
	~~$Q_F'=Q_F={l\0K}={K-1\02K}$ $~\then~Q_X'={K+1\0K}+1-Q_F={3K+3\02K}$.

\begin{center}
\begin{tabular}{||l|l|ll\TVR52|l||} \hline\hline
${16}_R$ & $|R,R,\sc,\sc\rangle$	&~ $Q_B\equiv{1\04}-{K\02}{l-k/2\0K}$
	& $\then ~~l\in{K-1\02}+2\IZ$		& $Q'_X=1-Q'_F$\\
${16}_{NS}$ & $|R,a,0,\sc\rangle$	&~ $Q_B\equiv l/2$	
	& $\then ~~l={K-1\02}\in 2\IZ$	& $Q'_X={K+1\02K}$\\
$\5{16}_R$ & $|R,R,s,\sc\rangle$	&~ $Q_B\equiv{3\04}-{K\02}{l-k/2\0K}$
	& $\then ~~l\in{K+1\02}+2\IZ$ 	& $Q'_X=2-Q'_F$ \\
$\5{16}_{NS}$ & $|R,c,0,\sc\rangle$	&~ $Q_B\equiv-l/2$	
	& $\then ~~l={K-1\02}\in 2\IZ$	& $Q'_X={3K+3\02K}$	\\
\hline\hline
\end{tabular}
\end{center}

{\VS-17	\tiny
\[ 1 + 4tT + 10{t^2}{T^2} + 20{t^3}{T^3} + 31{t^4}{T^4} + 
   40{t^5}{T^5} + 44{t^6}{T^6} + 40{t^7}{T^7} + 
   31{t^8}{T^8} + 20{t^9}{T^9} + 10{t^{10}}{T^{10}} + 
   4{t^{11}}{T^{11}} + {t^{12}}{T^{12}} + {T^{12}}\,x + 
   {t^4}\,{T^8}\,{x^2} + {t^8}\,{T^4}\,{x^3} + {t^{12}}\,{x^4}	\VS-27	
\]
}

\begin{center}
\begin{tabular}{||c|c|l|c\TVR74||} \hline\hline
${16}$ 	&  ~~~~ $\8{Q_B\equiv0}\atop n=n_0+s\,K$ ~~~~ & keep sign $+$ $=(-)^n$
	~\,	& 	$(J_s^{tot})^{2n}$ ~~$\bigl\{{0\le n<M~~~M\in2\IZ\atop
			0\le n < 2M~~M\not\in2\IZ}$ 
\\\hline\hline
R: $n\in K\IZ$	&	$l\in 2\IZ+{K-1\02} \atop\VR33	0\le l\le K-2$ 
		&	$(-)^n=(-)^s\s_X$
		&	${	Q'_X=1-l/K\atop	\VR43	\5Q'_X=1-l/K }$
\\\hline
R: $n\not\in K\IZ$ &  $n_0\in2\IZ+{K+1\02}\atop l=n_0-1, \5l=K-n_0-1$ 
		& 	$(-)^n=(-)^{n+1}\s_X$
		&  	${  Q'_X=1-{n_0-1\0K}\atop \VR43 \5Q'_X={n_0+1\0K}}$
\\\hline
NS: $n\in K\IZ$	&	$n'_0\neq0~~\then \atop n'_0-1=  l={K-1\02}\in2\IZ$
		& 	$(-)^n=(-)^n\s_X$
	  	&  	${Q'_X={K+1\02K}\atop \VR43 \5Q'_X={K+1\02K}}$
\\\hline
NS: $n'\in K\IZ$ & $n_0=l={K-1\02}\in2\IZ\atop	n'=n+{K+1\02}$
		& 	$(-)^n=(-)^{n'}\s_X$
		&  	${Q'_X={K+1\02K}\atop \VR43 \5Q'_X={K+3\02K}}$
\\\hline\hline\end{tabular}
\end{center}

\begin{center}
\begin{tabular}{||c|c|l|c\TVR74||} \hline\hline
$\5{16}$ 	& ~~~~ $\8{Q_B\equiv0}\atop n=n_0+s\,K$ ~~~~ 
		& 	keep sign $-$ $=(-)^{n+1}$
		& 	$(J_s^{tot})^{2n}$ ~~$\bigl\{{0\le n<M~~~M\in2\IZ\atop
			0\le n < 2M~~M\not\in2\IZ}$ 
\\\hline\hline
R: $n\in K\IZ$	&	$l\in 2\IZ+{K+1\02} \atop\VR33	0\le l\le K-2$ 
		&	$(-)^{n+1}=(-)^n\s_X$
		&	${	Q'_X=2-l/K\atop	\VR43	\5Q'_X=1-l/K }$
\\\hline
R: $n\not\in K\IZ$ &  $n_0\in2\IZ+{K-1\02}\atop l=n_0-1, \5l=K-n_0-1$ 
		& 	$(-)^{n+1}=(-)^{n+1}\s_X$
		&  	${  Q'_X=2-{n_0-1\0K}\atop \VR43 \5Q'_X={n_0+1\0K}}$
\\\hline
NS: $n\in K\IZ$	&	$n'_0\neq0~~\then \atop n'_0-1=  l={K-1\02}\in2\IZ$
		& 	$(-)^n=(-)^n\s_X$
	  	&  	${Q'_X={K+1\02K}\atop \VR43 \5Q'_X={K+1\02K}}$
\\\hline
NS: $n'\in K\IZ$ & $n_0=l={K-1\02}\in2\IZ\atop	n'=n+{K+1\02}$
		& 	$(-)^n=(-)^{n'}\s_X$
		&  	${Q'_X={K+1\02K}\atop \VR43 \5Q'_X={K+3\02K}}$
\\\hline\hline\end{tabular}
\end{center}

\del
\begin{center}
\begin{tabular}{||c|c|l|c\TVR74||} \hline\hline
$\5{16}$ 	& ~~~~~~ $Q_B\equiv0$ ~~~~~~	& ~~~ sign ~~~ 
		&	$(J_s^{tot})^{2n}$ :: $0\le n < M$ 
\\\hline\hline
R: $n\in K\IZ$	&	$l\in 2\IZ+{K-1\02} \atop\VR33	0\le l\le K-2$ 
		&	$1=+\s_X$
		&	${	Q'_X=2-l/K\atop	\VR43	\5Q'_X=2-l/K }$
\\\hline
R: $n\not\in K\IZ$ &  $n_0={K-1\02}\in2\IZ\atop n_0=n\mod K$ 
		& 	$(-)^n=(-)^{n+1}\s_X$
		&  	${  Q'_X=2-{n_0-1\0K}\atop \VR43 \5Q'_X=1+{n_0+1\0K}}$
\\\hline
NS: $n\in K\IZ$	&	$l={K-1\02}\in2\IZ	$	
		& 	$???=+\s_X$
	  	&  	${Q'_X={3K+3\02K}\atop \VR43 \5Q'_X={3K+1\02K}}$
\\\hline
NS: $n\not\in K\IZ$ & $n_0-1={K-1\02}\in2\IZ\atop n_0=n\mod K$
		& 	$???=(-)^{n+1}\s_X$
		&  	${Q'_X={3K+3\02K}\atop \VR43 \5Q'_X={3K-1\02K}}$
\\\hline\hline\end{tabular}
\end{center}
\enddel

\end{appendix}					\EndCount

%%%	\newpage

\bibliographystyle{ws-rv-van}			
\end{document}